\documentclass[final,3p,times,authoryear]{elsarticle}
\usepackage{amssymb}
\usepackage{rotating,longtable,boldline,amsmath}
\usepackage[header]{appendix}
\usepackage{comment}
\usepackage[normalem]{ulem}
\usepackage{xcolor}
\usepackage{soul}
\usepackage{mathrsfs}
\usepackage{amsbsy}
\usepackage{float}

\newcommand{\intwidth}[1]{\int_{-a/2}^{a/2}#1\:\mathrm{d}T}
\newcommand{\intlength}[1]{\int_{0}^{\ell}#1\:\mathrm{d}S}
\newcommand{\avgwidth}[1]{\left\langle #1\right\rangle_{T}}
\newcommand{\avglength}[1]{\left\langle #1\right\rangle_{S}} 

\newcommand{\br}{\mathbf{r}}
\newcommand{\bx}{\mathbf{x}}
\newcommand{\bd}{\mathbf{d}}
\newcommand{\be}{\mathbf{e}}
\newcommand{\bka}{\boldsymbol{\kappa}}

\newcommand{\cW}{\mathcal{W}}
\newcommand{\cF}{\mathcal{F}}
\newcommand{\cL}{\mathcal{L}}
\newcommand{\dab}{\delta_{\alpha\beta}}
\newcommand{\daap}{\delta_{\alpha\alpha'}}
\newcommand{\dbbp}{\delta_{\beta\beta'}}
\newcommand{\dapbp}{\delta_{\alpha'\beta'}} 
\newcommand{\das}{\delta_{\alpha S}}
\newcommand{\dbs}{\delta_{\beta S}}
\newcommand{\Aabapbp}{A_{\alpha\beta\alpha'\beta'}}
\newcommand{\Eab}{E_{\alpha\beta}}
\newcommand{\Eapbp}{E_{\alpha'\beta'}}
\newcommand{\Ecc}{E_{\gamma\gamma}}
\newcommand{\Ess}{E_{SS}}
\newcommand{\Est}{E_{ST}}
\newcommand{\Ett}{E_{TT}}
\newcommand{\nab}{n_{\alpha\beta}}

\newcommand{\nssref}{n_{SS}^{0}}
\newcommand{\Bab}{B_{\alpha\beta}}
\newcommand{\Bapbp}{B_{\alpha'\beta'}}
\newcommand{\Bcc}{B_{\gamma\gamma}}
\newcommand{\Bss}{B_{SS}}
\newcommand{\Bst}{B_{ST}}
\newcommand{\Btt}{B_{TT}}
\newcommand{\mab}{m_{\alpha\beta}}

\newcommand{\ed}{\epsilon^{\dag}}
\newcommand{\edd}{\epsilon^{\ddag}}

\newcommand \beq{\begin{equation}}
\newcommand \eeq{\end{equation}}
\newcommand \beqn{\begin{equation*}}
\newcommand \eeqn{\end{equation*}}
\newcommand \beqa{\begin{eqnarray}}
\newcommand \eeqa{\end{eqnarray}}
\newcommand \beqan{\begin{eqnarray*}}
\newcommand \eeqan{\end{eqnarray*}}

\newcommand{\pd}[2]{\frac{\partial #1}{\partial #2}} 
\journal{Journal of the Mechanics and Physics of Solids}

\begin{document}

\begin{frontmatter}



\title{Twisting instabilities in elastic ribbons with inhomogeneous pre-stress:\\
a macroscopic analog of %classical
thermodynamic phase transition}


\author[inst1]{Michael Gomez}

% \affiliation[inst1]{organization={Flexible Structures Laboratory, \'{E}cole Polytechnique F\'{e}d\'{e}rale de Lausanne (EPFL)}, %Department and Organization
%             addressline={Address One}, 
%             city={City One},
%             postcode={00000}, 
%             state={State One},
%             country={Country One}}

\author[inst1]{Pedro M. Reis}
\author[inst2]{Basile Audoly}

\address[inst1]{Flexible Structures Laboratory, Institute of Mechanical Engineering, \\ \'{E}cole Polytechnique F\'{e}d\'{e}rale de Lausanne (EPFL), 1015 Lausanne, Switzerland}
\address[inst2]{Laboratoire de M\'{e}canique des Solides, CNRS, Institut Polytechnique de Paris, 91120 Palaiseau, France}

% \affiliation[inst2]{organization={Department Two},%Department and Organization
%             addressline={Address Two}, 
%             city={City Two},
%             postcode={22222}, 
%             state={State Two},
%             country={Country Two}}

\begin{abstract}
We study elastic ribbons subject to large, tensile pre-stress confined to a central region within the cross-section. These ribbons can buckle spontaneously to form helical shapes, featuring regions of alternating chirality (phases) that are separated by so-called perversions (phase boundaries). This instability cannot be described by classical rod theory, which incorporates pre-stress through effective natural curvature and twist; these are both zero due to the mirror symmetry of the pre-stress. Using dimension reduction, we derive a one-dimensional (1D) `rod-like' model from a plate theory, which accounts for inhomogeneous pre-stress as well as finite rotations. The 1D model successfully captures the qualitative features of torsional buckling under a prescribed end-to-end displacement and rotation, including the co-existence of buckled phases possessing opposite twist, and is in good quantitative agreement with the results of numerical (finite-element) simulations and model experiments on elastomeric samples. Our model system provides a macroscopic analog of phase separation and pressure-volume-temperature state diagrams, as described by the classical thermodynamic theory of phase transitions.
\end{abstract}

% Our analysis highlights the competition between axial stretching and shear that underlies twisting and so provides insight into torsional instabilities observed in other prismatic solids.


%%\item Research highlight 1
%%\item Research highlight 2


\end{frontmatter}



\section{Introduction}
\label{sec:intro}

In many physical and biological settings, slender elastic filaments (rods or ribbons) possess internal structure: the material or geometric properties of the filament vary significantly over its cross-section, or non-uniform residual stresses arise due to active processes such as thermal expansion, swelling, or growth. For example, in biology, the tendrils of climbing plants coil upon contact with a support, yielding a spring-like attachment that supports the growing stem \citep{goriely1998,gerbode2012,wan2018}. At smaller scales, structured rods may arise from molecular assembly: the bacterial flagellar filament is formed by the polymerization of the protein flagellin into `protofilaments', whose conformations govern $11$ distinct helical forms \citep{calladine1975,kamiya1980}. In engineering, the microscopic structure of architectured materials can be designed \emph{a priori} to yield filaments with a desired global response. Examples include flexible porous strips whose drag coefficient may be tuned via the pattern of perforations \citep{guttag2018,jin2020,pezzulla2020}, and soft pneumatic actuators consisting of hollow elastomeric rods that coil upon inflation \citep{jones2021,becker2022}. 
% with non-uniform wall thicknesses, to grasp objects or act as artificial muscles in soft robotics    
    
Just as one curls a ribbon of cloth by irreversibly stretching its outer layer over a sharp blade \citep{prior2016}, structured filaments can exhibit dramatic shape changes that may be central to their function. This motivates understanding how the microscopic structure determines the global behavior. The canonical example is the bimetallic strip of Timoshenko \citep{timoshenko1925}, in which differential thermal expansion generates spontaneous curvature along the strip; engineering applications include clocks, thermostats, and circuit breakers \citep{wahl1944,timoshenko1961}. Many variations of this classic bilayer system have subsequently been studied. For example, differential swelling has been achieved in elastomeric bilayer shells \citep{pezzulla2016,pezzulla2018,lee2019}, photothermal hydrogels \citep{hauser2015}, polymer films via spatially-patterned crosslinking \citep{jamal2011,kim2012a,kim2012b}, and poroelastic sheets using directed fluid transport \citep{reyssat2011}. Other examples include differential hygroscopic expansion in pine cone scales and artificial bilayers \citep{dawson1997,reyssat2009,poppinga2018}, and artificial muscles based on differential Maxwell stresses in dielectric elastomers \citep{shian2015}; see also the reviews by \cite{chen2016} and \cite{vanManen2018}.

% exploding seed capsules driven by residual stresses \citep{hofhuis2016}; nanotubes fabricated from bilayers possessing different lattice constants \citep{schmidt2001}; soft walking robots whose locomotion is enabled by differential pneumatic inflation \citep{shepherd2011}

In this paper, we focus on elastic filaments whose complex, global shapes arise from non-uniform pre-stress. For such filaments, it is generally challenging to predict their global behavior based on the distribution of pre-stress, except in some special cases. For a slender filament subject to small pre-stress, it is well known that the system can be modeled as an Euler–Bernoulli rod with effective natural curvature and twist \citep{aharoni2012,moulton2020}. Indeed, several studies have shown convergence (in a rigorous mathematical sense) of three-dimensional (3D) elasticity to a one-dimensional (1D) rod-like model, whose precise form depends on the assumed scalings for the elastic energy and external loading \citep{kupferman2014,freddi2016,cicalese2017,agostiniani2017,kohn2018}.

For large pre-stress, however, the filament can undergo an elastic instability on the scale of its cross-section dimensions. Thus, the kinematic assumptions underlying classical rod models do not apply. A notable example is an elastic bi-strip formed by gluing a uniaxially pre-stretched strip to a second strip that is initially unstressed \citep{huang2012,liu2014}; the resulting distribution of pre-stress resembles a step function. This configuration is a variation of the classic bilayer system in which the strip's thickness is comparable to its width, so that complex bending and twisting deformations are observed. In particular, depending sensitively on the cross-section geometry and the magnitude of the pre-strain, \cite{huang2012} and \cite{liu2014} demonstrated that the bi-strip either buckles globally to adopt a helical shape, or it forms a `hemihelix' characterized by a series of helices with alternating chiralities. As discussed by \cite{lestringant2017}, Euler-Bernoulli rod theory cannot predict the wavelength selection of the experimental hemihelical shapes, since the theory fails to capture the cross-section deformations associated with this small-wavelength instability. The goal of the present work is to derive a 1D model that can capture helical buckling in a similar, albeit simpler, system, by applying dimension reduction to a more general modeling framework that does not make \emph{ad hoc} kinematic assumptions about how cross-sections deform.

% Instead, \cite{lestringant2017} appealed to a hierarchy of models ranging from a double-beam model to a 3D hyperelastic prismatic solid.
% In situations where classical rod models do not apply, it may still be desirable to describe such systems using reduced models. It is then necessary to use a more general modeling framework, which does not assume small pre-stress or make \emph{ad hoc} kinematic assumptions about how cross-sections deform.

Broadly, the aim of dimension reduction methods, starting from a general description of an elastic continuum, is to systematically exploit the slenderness of the structure (in one or more dimensions) to derive a lower-dimension model. Usually, this procedure yields an energy density along an effective centerline or mid-surface. The resulting models have the benefit of being simpler to analyze while retaining the salient features of the full system; such features include multi-stability of equilibrium states and their associated bifurcations. These methods have been applied successfully to a wide range of mechanical systems, including tensile necking in prismatic solids \citep{coleman1988,audoly2016}, elastocapillary necking of cylindrical gels \citep{lestringant2020b}, bulges in hyperelastic membranes \citep{lestringant2018,yu2023}, and morpho-elastic rods \citep{lessinnes2017,moulton2020,kaczmarski2022}. 
    
Ribbon models are at the forefront of dimension-reduction methods for slender structures. Unlike an elastic rod --- whose thickness and width are of comparable size and much smaller than the length --- a ribbon is characterized by a `flattened' cross-section with both small thickness-to-width and width-to-length ratios. As a result, ribbons exhibit mechanical behavior between that of a plate and a rod \citep{levin2021}: as well as undergoing large, global displacements akin to a rod, their extended width leads to a strong coupling between geometry and mechanics, in which Poisson effects, isometric transformations, and stress localization may play a key role. Building on early work by \cite{sadowsky1930} and \cite{wunderlich1962} on uniform, rectangular ribbons undergoing inextensible deformations, recent studies have considered the effects of variable width, natural curvature, mid-surface extensibility, and strain gradients \citep{starostin2015,dias2015,efrati2015,taffetani2019,brunetti2020,audoly2021b,levin2021,kumar2023}.
%neukirch2021,
     
Here, we study pre-stressed ribbons as a model system to develop and test reduced-dimension models when classical theories do not apply. Specifically, we consider the symmetric version of the bi-strip studied by \cite{huang2012} and \cite{liu2014}: an elastic ribbon formed by bonding a pre-stretched strip to \emph{two identical} strips that are initially unstressed. This system, referred to as a `\textit{trilayer}', has been briefly discussed by \cite{kohn2018} in the thin-rod limit (i.e., for a comparable thickness and width), who noted that the effective curvature and twist are zero due to the mirror symmetry of the pre-stress; the same conclusion holds for ribbons \citep{freddi2016}. Nevertheless, twisting instabilities \emph{have} been observed in systems where a similar distribution of pre-stress arises, including the ruffled blades of kelp \citep{koehl2008}, baromorph elastomers \citep{siefert2019} and patterned fabric sheets \citep{gao2020,siefert2020a}. The common feature of these systems is that the pre-stress is large; in being limited to small pre-stress, the classical theory evidently cannot capture these kind of twisting instabilities. We note that large residual stress has been incorporated into a 1D morpho-elastic theory by \cite{moulton2020}, though their analysis is limited to the thin-rod limit. We will derive a 1D model for ribbons that carefully accounts for large, inhomogeneous pre-stress, by extending the extensible ribbon model recently developed by \cite{audoly2021b}. While the twist has a single preferred value in classical theories, we demonstrate that the pre-stress triggers an instability associated with two competing values, which govern a nonlinear effective behavior not covered by existing reduction methods.
    
Our manuscript is organized as follows. In \S\ref{sec:problemdefn}, we define the ribbon system under consideration, including the two loading scenarios (end-shortening and end-rotation) investigated. In \S\ref{sec:ribbonmodel}, we derive the reduced ribbon model and discuss its analogy with the classical theory of thermodynamic phase transitions. We then discuss the numerical and experimental methods we used to quantify the behavior of pre-stressed ribbons in \S\ref{sec:methods}. In \S\ref{sec:results}, we present results of loading by end-shortening and end-rotation. In each scenario, we compare the results of numerical simulations, experiments on elastomeric ribbons, and the predictions of the ribbon model. In \S\ref{sec:ribbonvalidity}, we consider the limitations of the ribbon model. Finally, in \S\ref{sec:discussionconclusions}, we discuss our findings and conclude.

\section{Definition of the problem}
\label{sec:problemdefn}

\begin{figure}
\centering
\includegraphics[width=\textwidth]{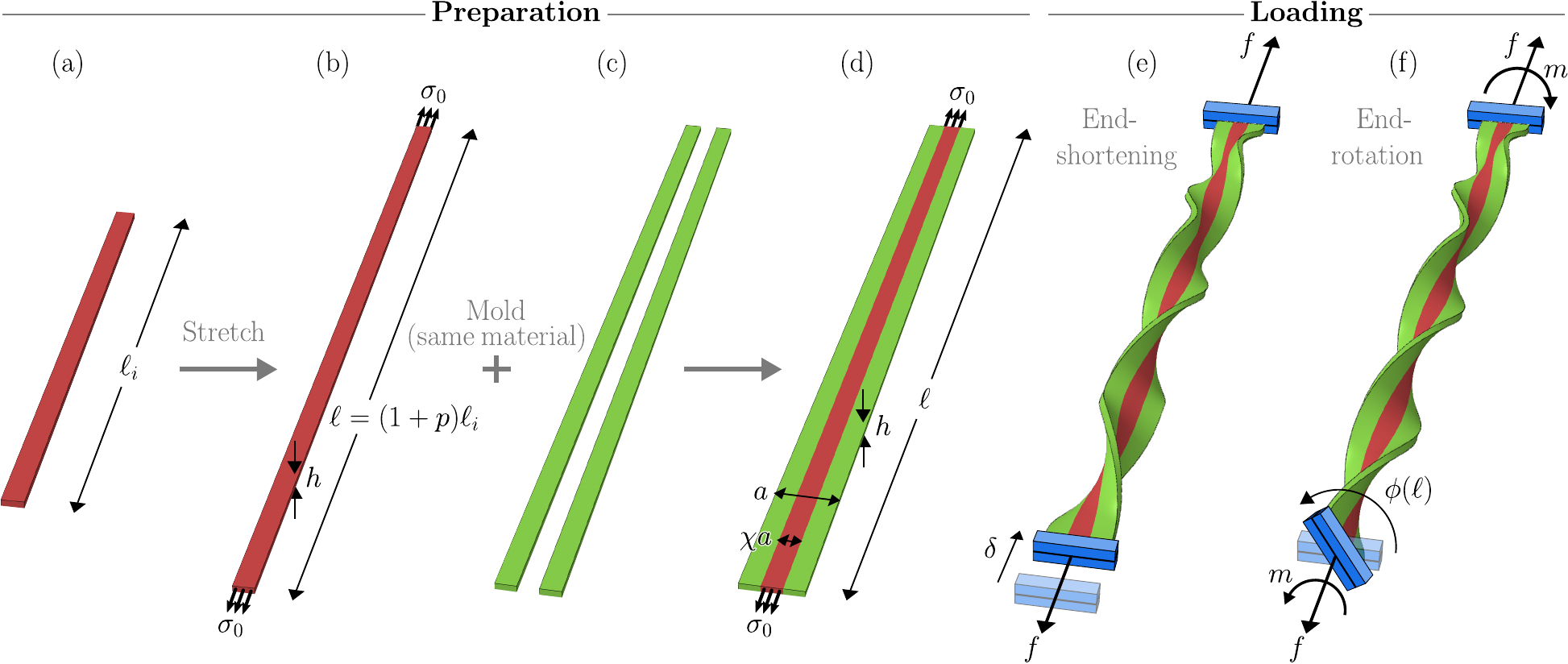} 
\caption{Definition of the problem: preparation of an elastic ribbon with inhomogeneous, uniaxial pre-stress and the loading scenarios we consider. (a)--(b) An elastic strip with rectangular cross-section and length $\ell_i$ is subject to a uniaxial strain $p \geq 0$, increasing its length to $\ell=(1+p)\ell_i$. (c)~The stretched `inner' strip (red) is bonded along its outer edges to two identical `outer' strips (green), composed of the same material as the inner strip. (d) The bonded strips form a ribbon with uniform geometry (the `fully-unrelaxed' configuration) in which the pre-stress is non-zero only in the central, red region. (e) With its ends clamped parallel to the planar state, the fully-unrelaxed ribbon is subjected to an end-shortening $\delta$ by displacing one clamp along the ribbon axis, causing torsional buckling. (f) The end-shortened ribbon is further loaded by an end-rotation $\phi(\ell)$.}
\label{fig:problemdefn}
\end{figure}

We investigate the model system shown schematically in Fig.~\ref{fig:problemdefn}: an elastic ribbon with a discontinuous distribution of pre-stress in the cross-section, prepared as follows. Starting with an elastic strip with a uniform and rectangular cross-section (Fig.~\ref{fig:problemdefn}a), we apply a uniaxial strain $p \geq 0$ so that its length becomes multiplied by a factor $(1+p)$ (Fig.~\ref{fig:problemdefn}b); the corresponding stress is $\sigma_0$. The stretched strip is subsequently bonded along its outer, lateral edges to two identical strips (Fig.~\ref{fig:problemdefn}c). Composed of the same material, the thickness and length of these additional strips exactly match those of the pre-stretched strip when they are unstressed (Fig.~\ref{fig:problemdefn}d). Hence, after bonding, the strips form a ribbon of (constant) thickness $h$, width $a$, and length $\ell$, in which the pre-stress is non-zero (and tensile) only in an `inner' region of width $\chi a$; here $\chi \in (0,1)$ is the dimensionless area fraction. The inner region lies symmetrically in the cross-section so that the overall distribution of pre-stress is also symmetric. We refer to the configuration in Fig.~\ref{fig:problemdefn}d --- in which the outer region (green) is unstressed --- as the `fully-unrelaxed' shape. Furthermore, we refer to the system as a \emph{ribbon} rather than a \emph{rod} or a \emph{plate}, owing to its small thickness-to-width and width-to-length ratios: throughout, we consider $h \ll a \ll \ell$. 
   
We will explore the mechanical behavior of the ribbon under two loading scenarios.
\begin{itemize}
\item Firstly, starting in the fully-unrelaxed configuration, we allow its ends to shorten by a distance $\delta > 0$ (referred to as the \emph{end-shortening}) while simultaneously clamping the ends to prevent transverse displacement and rotation; see Fig.~\ref{fig:problemdefn}e. The end-shortening effectively controls the \emph{average axial strain} in the ribbon, which serves as a control parameter. In general, the ribbon remains planar for small end-shortenings: with symmetric pre-stress, the ribbon cannot relax the pre-stress by developing spontaneous curvature (as occurs with the classical bimetallic strip). Instead, we will show that, above a critical end-shortening, the ribbon undergoes torsional buckling: it deforms significantly out of plane, forming helicoidal-like shapes in which the cross-section twists about the ribbon centerline at a well-defined spatial rate. Broadly speaking, twisting allows the ribbon to relieve some compressive stress in the outer region while the inner region remains under tension.
\item In the second loading scenario (Fig.~\ref{fig:problemdefn}f), we rotate one clamp by an \emph{angle $\phi(\ell)$ about the ribbon axis} (which serves as an alternate control parameter) while fixing the end-shortening.
\end{itemize}
In both loading scenarios, we consider the axial force, $f$, and moment, $m$, exerted on the clamps.
% (as suggested by the notation, we will later use $\phi(\cdot)$ to denote the rotation angle of each cross-section in the ribbon)

We seek to understand when torsional buckling occurs and how the resulting twisting influences the overall, macroscopic behavior of the ribbon. Specifically, using a 1D `rod-like' model and simulations based on the finite element method (FEM), we characterize the local twist rate (i.e., the rate at which cross-sections rotate about the ribbon centerline) as a function of the applied end-shortening $\delta$, end-rotation $\phi(\ell)$, pre-strain $p$ and area fraction $\chi$. We quantify the macroscopic behavior of the ribbon using the axial force, $f$, and moment, $m$ (Figs.~\ref{fig:problemdefn}e--f). We will compare our analytical and numerical results with our experiments on elastomeric samples.
%, for which the force and moment can readily be measured.

Furthermore, a key feature of the torsional buckling exhibited by our ribbons is that the helicoidal shapes generally exist in regions of alternating chirality (Figs.~\ref{fig:problemdefn}e--f). These regions can be regarded as distinct thermodynamic phases in which the twist rate acts as an order parameter. The phase boundaries between neighboring regions, where the twist rate rapidly changes sign according to the chiralities of the regions, are similar to the \emph{perversions} observed in elastic rods with intrinsic curvature \citep{goriely1998,mcmillen2002}. Following \cite{huang2012}, we also refer to them here as perversions, even though the ribbon has zero intrinsic curvature. The torsional buckling can then be viewed as a phase separation process, in which the planar, homogeneous state becomes unstable to distinct buckled phases that co-exist in equilibrium. In addition to developing a quantitative understanding of the instability, we aim to substantiate this thermodynamic analogy through the analytical insight afforded by the 1D model.

\section{Ribbon model of inhomogeneously pre-stressed ribbons}
\label{sec:ribbonmodel}

In this section, we develop a theoretical model for an elastic ribbon subject to inhomogeneous pre-stress (as shown in Fig.~\ref{fig:problemdefn}). In particular, we seek a simplified model to gain analytical understanding of the torsional instability. To understand the behavior beyond the onset of buckling, our model needs to account for extensibility of the ribbon mid-surface and finite rotations of its cross-section with respect to the laboratory frame. While a geometrically-exact shell theory may be an obvious candidate, the complexity of the governing equations means that analytical progress is generally not possible. Recently, \cite{audoly2021b} have demonstrated that an extensible ribbon model, accounting for finite rotations, can be systematically derived from a geometrically-nonlinear plate model, using a dimension reduction that is asymptotically valid in the slenderness limit characteristic of ribbons. The result is a 1D `rod-like' model, in which the strain energy is expressed solely in terms of `macroscopic' strains (stretching, bending, twisting) attached to the ribbon centerline.

% Moreover, the commonly used F\"{o}ppl-von-K\'{a}rm\'{a}n plate model can only account for moderate rotations in the laboratory frame, and is generally not analytically tractable.

We adapt the extensible ribbon model of \cite{audoly2021b} to incorporate inhomogeneous, uniaxial pre-stress. Because our derivation up to \S\ref{sec:dimredresult} closely follows that proposed by these authors, we do not present all details of the dimension reduction; instead, we outline the main steps of the method and highlight the elements that are different here. Throughout, we will focus on steady deformations, i.e., we suppose that the end-shortening and end-rotation are applied quasi-statically. For simplicity, we assume that both inner and outer regions (Fig.~\ref{fig:problemdefn}d) are composed of the same isotropic, homogeneous material. In addition to the slenderness assumption $h \ll a \ll \ell$, we assume a small pre-strain, $p \ll 1$, as well as small strains during end-shortening and end-rotation; this requires $\delta/\ell \ll 1$ and $a\phi(\ell)/\ell \ll 1$ (see Figs.~\ref{fig:problemdefn}e--f). Hence, a linearly elastic constitutive law is appropriate, with Young's modulus $Y$ and Poisson ratio $\nu$. Furthermore, because we are primarily interested in a twisting instability, for which the centerline remains approximately straight (Figs.~\ref{fig:problemdefn}e--f), we limit attention to the case of combined stretching and twisting: we only seek solutions with straight centerline (zero bending strains). Together, these assumptions allow us to make significant analytical progress. Later, we will compare our predictions with numerical simulations and experiments for which the strains are not necessarily small, demonstrating good quantitative agreement despite the ribbon model being formally valid only for small strains.

% and was successfully applied by \cite{audoly2021b} to several buckling problems, namely lateral-torsional buckling of a ribbon under gravity, classical Euler buckling under axial compression, and buckling of a twisted ribbon under tension.

% \footnote{While it is possible to adapt the general model developed by \cite{audoly2021b} (with centerline bending) to incorporate inhomogeneous pre-stress, the resulting 1D model is extremely cumbersome, even for relatively simple distributions of pre-stress like that considered in this paper. In particular, it leads to lengthy equilibrium equations for the macroscopic bending and twisting strains that are not particularly informative. As we will show, the assumption of zero centerline bending significantly simplifies the model while retaining the key ingredients underlying the torsional instability. Moreover, it is exactly satisfied throughout a large region of parameter space.}.

\subsection{Kinematic description of the ribbon}
\label{sec:ribbonkinematics}
\begin{figure}
\centering
\includegraphics[width=\textwidth]{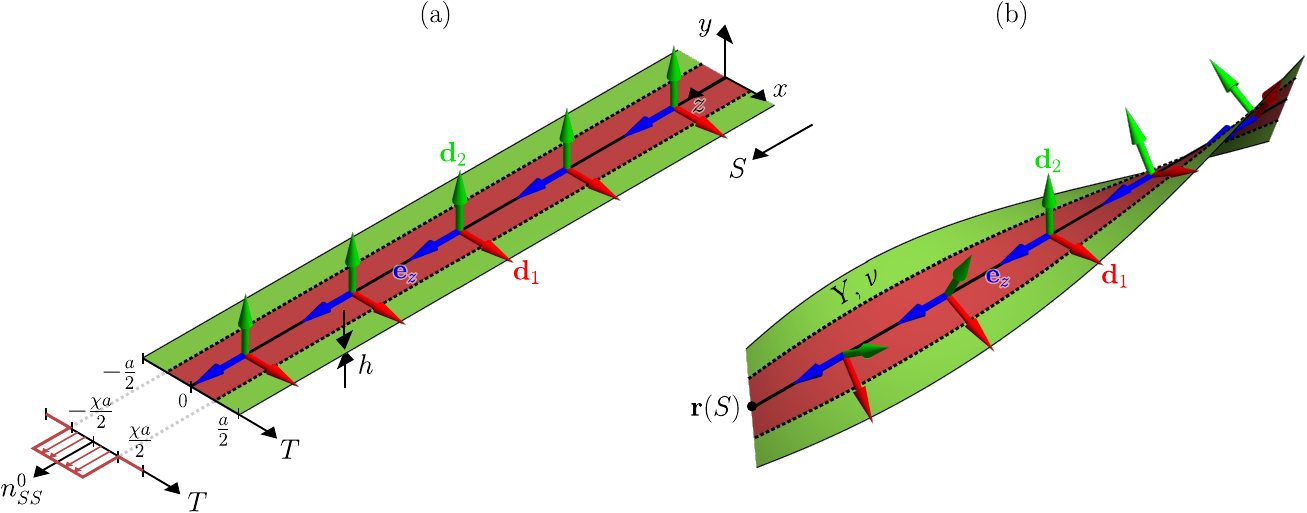} 
\caption{The centerline-based kinematic description used in the ribbon model developed in \S\ref{sec:ribbonmodel}. (a) Reference configuration, corresponding to the fully-unrelaxed configuration shown in Fig.~\ref{fig:problemdefn}d. The discontinuous distribution of pre-stress resulting from the fabrication procedure in Figs.~\ref{fig:problemdefn}a--d is also drawn. (b) Deformed configuration. (In both panels only a portion of the total length is drawn.)}
\label{fig:ribbonschematic}
\end{figure}

To facilitate the dimension reduction, we describe the ribbon kinematics using a centerline and a set of orthogonal directors (Fig.~\ref{fig:ribbonschematic}). Throughout this section, we take our reference state to be the fully-unrelaxed configuration shown in Figs.~\ref{fig:problemdefn}d and \ref{fig:ribbonschematic}a, i.e., we define displacements and strains relative to this configuration. We take Cartesian coordinates in the laboratory frame such that the reference mid-surface lies in the $(x,z)$ plane, with the $z$-axis parallel to the ribbon centerline (Fig.~\ref{fig:ribbonschematic}a); the ribbon ends correspond to $z = 0$ and $z = \ell$. The corresponding unit vectors in Cartesian coordinates are $\lbrace\be_x,\be_y,\be_z\rbrace$. We also define material coordinates $(S,T)$ along the longitudinal and transverse directions, respectively, where $0 \leq S \leq \ell$ and $-a/2 \leq T \leq a/2$ (thus, in the reference configuration, we can identify $S = z$ and $T = x$). The reference configuration has a resultant pre-stress (defined as the bulk stress integrated over thickness) that is uniaxial in the longitudinal direction, denoted $\nssref(T)$. This has the form (Fig.~\ref{fig:ribbonschematic}a):
\beq
\nssref(T) = \begin{cases} h\sigma_0 \quad & \mathrm{if} \quad |T| < \chi a/2, \\ 0  \quad & \mathrm{if} \quad \chi a/2 < |T| < a/2, \end{cases} \label{eqn:prestress}
\eeq
where, under the assumption of linear elasticity, $\sigma_0 = Y p$.

In the deformed configuration, we denote the centerline of the ribbon (defined to be the centroid of the cross-section for each $S$) by $\br(S)$; see Fig.~\ref{fig:ribbonschematic}b. Under the assumption of zero centerline bending, the tangent vector is everywhere parallel to the $z$-axis:
\beq
\br'(S) = \left[1 + \epsilon(S)\right]\be_z,
\eeq
where $\epsilon(S)$ is the macroscopic axial strain relative to the reference configuration, controlled (indirectly) by the imposed end-shortening.

To complete the description of the local orientation of the ribbon, we introduce the vectors $\bd_1(S)$ and $\bd_2(S)$ (referred to as \emph{directors} in the Kirchhoff theory of rods) such that $\lbrace\bd_1,\bd_2,\be_z\rbrace$ forms a right-handed orthonormal basis for each $S$. In the absence of centerline bending, $\bd_1$ and $\bd_2$ are simply a rotation of the Cartesian unit vectors $\be_x$ and $\be_y$, respectively, about the $z$-axis (Fig.~\ref{fig:ribbonschematic}). Denoting the angle of rotation by $\phi(S)$, we have
\beq
\bd_1 = \cos\phi(S)\,\be_x + \sin\phi(S)\,\be_y, \quad \bd_2 = \be_z \times \bd_1 = -\sin\phi(S)\,\be_x + \cos\phi(S)\,\be_y. \label{eqn:directorsd1d2}
\eeq
The twisting strain (spatial rate of twist), denoted $k(S)$, is
\beq
k(S) = \phi'(S).
\eeq
% As shown in Fig.~\ref{fig:ribbonschematic}, the vectors $\bd_1$ and $\bd_2$ span the cross-section perpendicular to $\be_z$, changing orientation as the ribbon bends and twists. In the reference configuration, $(\bd_1,\bd_2) = (\be_x,\be_y)$ (Fig.~\ref{fig:ribbonschematic}a). 

% While the above describes the average position and orientation of each cross-section in the deformed ribbon, we still need 
To characterize how the ribbon deforms within each cross-section, we introduce the displacement components in the director basis, $(u,v,w)$, such that the deformed position of the point with material coordinates $(S,T)$ is
\beq
\bx(S,T) = \br(S) + \left[T+u(S,T)\right]\bd_1(S) + w(S,T)\,\bd_2(S) + v(S,T)\,\be_z(S). \label{eqn:kinematiccurrentconfig}
\eeq
The `macroscopic' strains $\epsilon$ and $k$ --- characterizing the average deformation of each cross-section --- together with the `microscopic' displacements $(u,v,w)$ complete the kinematic description of the ribbon. It is also necessary to include the following kinematic constraints. Because we define the centerline to be the centroid of each cross-section, we must impose that the displacements $(u,v,w)$ have zero mean over the interval $T \in (-a/2,a/2)$:
\beq
\intwidth{u(S,T)} = 0, \quad \intwidth{v(S,T)} = 0, \quad \intwidth{w(S,T)} = 0, \qquad S \in (0,\ell). \label{eqn:kinematiccentroid}
\eeq
An additional constraint is needed to remove any indeterminacy in the definition of $\bd_1$ and $\bd_2$; from Eq.~\eqref{eqn:directorsd1d2}, this is equivalent to uniquely specifying the twist angle $\phi(S)$ for each $S$. We require that $\phi(S)$ indeed corresponds to an `average' twist angle of the cross-section about the $z$-axis, in the sense that the first moment of the out-of-plane displacement $w$ is zero:
\beq
\intwidth{T w(S,T)} = 0, \qquad S \in (0,\ell). \label{eqn:kinematictwistangle}
\eeq

Before proceeding, we determine the orders of magnitude of the macroscopic strains in terms of the ribbon parameters. These scalings inform which terms to retain in a weakly-nonlinear plate model of the ribbon, before any dimension reduction is applied. Following \cite{audoly2021b}, we perform a scaling analysis in the limit $h \ll a$ under consideration, assuming that (i) the stretching and twisting contributions to the elastic energy all balance at leading order; and (ii) the in-plane strain can be estimated from twisting over a lengthscale comparable to the plate width $a$. This yields\footnote{The scalings in Eq.~\eqref{eqn:scalingforms} are equivalent to those in \cite{audoly2021b}, up to coefficients $\sqrt{12\left(1-\nu^2\right)}$ of order 1.}
\beq
\epsilon = O\left(\frac{h^2}{a^2}\right), \quad k = O\left(\frac{h}{a^2}\right). \label{eqn:scalingforms}
\eeq

\subsection{Elastic energy of homogeneous solutions}
\label{sec:ribbonenergyformulation}

% Using the scaling behavior \eqref{eqn:scalingforms}, the dominant contributions to the strains arising due to deformation, $\Eabr$, in the limit $h \ll a$ are given by \citep{audoly2021b}
% \beqa
% \Essr(S,T) & = & \epsilon(S) + \frac{\left[k(S)\right]^2}{2}T^2 + k(S)T\pd{w}{S} + \pd{v}{S} + \frac{1}{2}\left(\pd{w}{S}\right)^2, \nonumber \\
% \Estr(S,T) & = & -\frac{k(S)}{2}\left[w(S,T) - T\pd{w}{T}\right] + \frac{1}{2}\left(\pd{u}{S} + \pd{v}{T} + \pd{w}{S}\pd{w}{T}\right), \nonumber \\
% \Ettr(S,T) & = & \pd{u}{T} + \frac{1}{2}\left(\pd{w}{T}\right)^2. \label{eqn:membranestrainsdefm}
% \eeqa
% \beqa
% \Bssr(S,T) & = & k'(S)T + \pdd{w}{S}, \nonumber \\
% \Bstr(S,T) & = & k(S) + \pddm{w}{S}{T}, \nonumber \\
% \Bttr(S,T) & = & \pdd{w}{T}. \label{eqn:bendingstrainsdefm}
% \eeqa

We denote the in-plane (membrane) strain components by $\Eab$ and the bending strains by $\Bab$; here and throughout, Greek indices are limited to the in-plane directions, $\alpha,\beta \in \lbrace S,T\rbrace$. By first calculating the deformation gradient associated with $\bx(S,T)$ in Eq.~\eqref{eqn:kinematiccurrentconfig}, then expanding terms using the scalings in Eq.~\eqref{eqn:scalingforms}, the dominant contributions to the strains in the limit $h \ll a$ were calculated by \cite{audoly2021b}. These strains are geometrically nonlinear and account for gradients in both the macroscopic strains $(\epsilon,k)$ and microscopic displacements $(u,v,w)$ along the ribbon centerline. In our following analysis, we neglect gradient effects along the centerline by limiting attention to `homogeneous solutions', independent of the longitudinal coordinate $S$:
\beq
(\epsilon,\ k) = \mathrm{constant}, \qquad u = u(T), \quad v = v(T), \quad w = w(T). \label{eqn:homogansatz}
\eeq
Given the separation of lengthscales $a \ll \ell$ inherent to the ribbon geometry, this simplification is asymptotically valid when $(\epsilon,k)$ are no longer independent of $S$ but vary on a lengthscale much larger than the ribbon width $a$. More precisely, the minimizers of the resulting elastic energy (subject to appropriate boundary conditions) are the dominant contribution when the solution of the full system is expanded in the limit $h \ll a \ll \ell$ \citep{hodges2006,audoly2021a}. The gradient terms involving $S$-derivatives of $(\epsilon,k)$ only enter the expansion at higher order.
%lestringant2020a,

Considering homogeneous solutions of the form \eqref{eqn:homogansatz} with zero centerline bending, the strain components simplify to (primes now denoting $T$ derivatives)
\beqa
&& \Ess(T) = \epsilon + \frac{k^2}{2}T^2, \qquad 
\Est(T) = -\frac{k}{2}\left[w(T) - T w'(T)\right] + \frac{1}{2}v'(T), \qquad
\Ett(T) = u'(T) + \frac{1}{2}\left[w'(T)\right]^2, \nonumber \\
&& \Bss(T) = 0, \qquad
\Bst(T) = k, \qquad
\Btt(T) = w''(T). \label{eqn:strainsdefmhomogeneous}
\eeqa
In the absence of the macroscopic strains $\epsilon$ and $k$, the above equations reduce to the usual (weakly-nonlinear) von K\'{a}rm\'{a}n strain-displacement relations for plates, as expressed in a Cartesian frame. Note that the longitudinal strain $\Ess$ is the macroscopic strain $\epsilon$ plus a contribution $(k\,T)^2/2$ depending on the transverse coordinate $T$, which captures the fact that, in the presence of twisting, fibers initially parallel to the centerline are deformed into helices with helical radius $|T|$. Thus, twisting induces longitudinal extension $E_{SS}$ that is more pronounced on the edges of the ribbon, where $T^2$ is larger. This explains why helical buckling can effectively relax compressive pre-stress that is localized along the edges of the ribbon.

Owing to the small thickness $h \ll a, \ell$, the elastic energy and constitutive relations are now calculated in the framework of classical thin-plate theory. The total strain energy associated with homogeneous solutions is $\ell \cW$, where $\cW$ is the strain energy per unit length. In the presence of uniaxial pre-stress $\nssref(T)$, we can write $\cW$ as (using Einstein notation for summation over repeated indices)
\beq
\cW = \frac{1}{2}\intwidth{\left[\Eab(T)\Aabapbp\Eapbp(T) + \frac{h^2}{12}\Bab(T)\Aabapbp\Bapbp(T)\right]} + \intwidth{\nssref(T)\Ess(T)}, \label{eqn:strainenergy}
\eeq
where $\Aabapbp$ is the stiffness tensor ($\dab$ is the Kronecker delta):
\beqn
\Aabapbp \equiv \frac{Y h}{1-\nu^2}\left[(1-\nu)\daap\dbbp + \nu\dab\dapbp\right].
\eeqn
The constitutive laws for the resultant membrane stresses, $\nab$, are then derived from Eq.~\eqref{eqn:strainenergy} as
\beqa
\nab(T) % & = & \pd{\cW}{\Eab} \nonumber \\
& = & \Aabapbp\Eapbp(T) + \nssref(T) \das\dbs % \nonumber \\ & = &
= \frac{Y h}{1-\nu^2}\left[(1-\nu)\Eab(T) + \nu\Ecc(T)\dab\right] + \nssref(T) \das\dbs, \label{eqn:constvmembrane}
\eeqa
and similarly for the resultant bending stresses, $\mab$:
\beqa
\mab(T) % & = & \pd{\cW}{\Bab} \nonumber \\
& =  &\frac{h^2}{12}\Aabapbp\Bapbp(T) % \nonumber \\ & = &
= \frac{Y h^3}{12\left(1-\nu^2\right)}\left[(1-\nu)\Bab(T) + \nu\Bcc(T)\dab\right], \qquad \alpha,\beta \in \lbrace S,T\rbrace. \label{eqn:constvbending}
\eeqa
These expressions resemble the standard constitutive relations for thin (non pre-stressed) plates, with an additional term arising from the uniaxial pre-stress $\nssref$.

\subsection{Dimension reduction via relaxation of microscopic displacements}
\label{sec:dimredresult}
Substituting the expressions in Eq.~\eqref{eqn:strainsdefmhomogeneous} for the strain components, the strain energy in Eq.~\eqref{eqn:strainenergy} can be written as a functional of the microscopic displacements and macroscopic strains:
\beqn
\cW = \cW\left(u,v,w;\epsilon,k\right).
\eeqn
For each combination of (constant) strains $(\epsilon,k)$, the displacements $\left(u(T),v(T),w(T)\right)$ are determined by the condition that $\cW$ is stationary subject to the kinematic constraints in Eqs.~\eqref{eqn:kinematiccentroid}--\eqref{eqn:kinematictwistangle}. This `relaxation' procedure allows us to eliminate $\left(u,v,w\right)$ in favor of $(\epsilon,k)$, leading to a 1D rod-like theory with energy density $\cW=\cW(\epsilon,k)$ along the ribbon centerline.

The relaxation procedure is presented in \ref{sec:appendixrelaxation}. The key result is that, as in the case without pre-stress considered by \cite{audoly2021b}, it is possible to reduce the Euler-Lagrange equations for $\left(u(T),v(T),w(T)\right)$ to a fourth-order boundary-value problem for the out-of-plane component, $w(T)$. The solution is unique and is given by $w(T) = 0$ for all $T$: with zero centerline bending, cross-sections rotate about the straight centerline and remain straight. Note that they can still stretch or contract by a Poisson effect, as $u(T) \neq 0$ in general. The solution is considerably more involved when centerline bending or strain gradients are considered.

After applying the relaxation procedure, the strain energy in Eq.~\eqref{eqn:strainenergy} reduces to (see \ref{sec:appendixrelaxation})
\beq
\cW\left(\epsilon,k\right) = \frac{Y h a}{2}\left\lbrace\left(\epsilon-\ed\right)^2 -{\ed}^2 + \frac{a^2}{12}\left[\frac{2 h^2}{a^2(1+\nu)} + (\epsilon-\ed) + \edd \right]k^2 + \frac{a^4}{320}k^4\right\rbrace. \label{eqn:ribbonenergydim}
\eeq
Here $\ed$ and $\edd$ are (constant) strains that depend on the distribution of pre-stress in Eq.~\eqref{eqn:prestress}:
\beq
\ed \equiv -\frac{1}{Yh}\avgwidth{\nssref} = -\chi p, \quad \edd \equiv \frac{12}{Yha^2}\avgwidth{T^2\left(\nssref-\avgwidth{\nssref}\right)} = -\chi\left(1-\chi^2\right)p, \label{eqn:epsilondags}
\eeq
where here, and in later equations, $\avgwidth{\cdot}$ and $\avglength{\cdot}$ denote averages over the ribbon width and length, respectively:
\beqn
\avgwidth{\cdot} \equiv \frac{1}{a}\intwidth{(\cdot)(T)}, \quad \avglength{\cdot} \equiv \frac{1}{\ell}\intlength{(\cdot)(S)}.
\eeqn

%Because we did not incorporate gradient terms in the strain energy, there are no terms in $\epsilon'(S)$, $k'(S)$, $\epsilon''(S)$, $k''(S)$ etc.~in Eq.~\eqref{eqn:ribbonenergydim} --- the energy, and its corresponding equilibrium solutions, applies locally (more specifically, on a lengthscale comparable to the plate width $a$) for which $\epsilon$ and $k$ are approximately constant. Nevertheless,

As discussed in \S\ref{sec:problemdefn}, we subject the ribbon to \emph{global} constraints via the end-shortening $\delta$ and end-rotation $\phi(\ell)$. These constraints impose, respectively, an average axial strain $-\delta/\ell$ and twist strain $\phi(\ell)$ along the ribbon centerline, $S \in (0,\ell)$. Therefore, we have
\beq
\avglength{\epsilon} = -\frac{\delta}{\ell}, \quad \avglength{k} = \frac{\phi(\ell)}{\ell}. \label{eqn:avgekdim}
\eeq

\subsection{Interpretation of the pre-stress coefficients $\ed$ and $\edd$}
To interpret the quantity $\ed$, we consider the longitudinal force (tension) in the absence of twisting ($k=0$):
\beqn
N_0(\epsilon) = \pd{\cW}{\epsilon}(\epsilon,0)=Y h a\left(\epsilon-\ed\right).
\eeqn
In the fully-unrelaxed configuration, the longitudinal force is tensile, $N_0(0)=-Y h a \ed>0$, as $\ed<0$ by Eq.~\eqref{eqn:epsilondags}. We also have $N_0(\ed)=0$, implying that $\ed$ is the (negative, hence contractile) overall strain at which the tension $N_0(\epsilon)$ goes from positive to negative when the ribbon is allowed to shorten.  For a very long ribbon, $\ell\gg a$, the condition of Euler buckling is precisely that the longitudinal force goes through zero; the buckling threshold is therefore
\beq
\epsilon = \ed \quad \textrm{(onset of Euler buckling),} \label{eq:OnsetOfEulerBuckling}
\eeq
 recalling that $\ed<0$. This is consistent with the fact that $\ed$ is defined in Eq.~\eqref{eqn:epsilondags} as proportional to the \emph{average} pre-stress.

Next, we consider the term in square brackets in Eq.~\eqref{eqn:ribbonenergydim}. Appearing in a factor of $k^2$, it can be interpreted as an incremental (scaled) twisting modulus, and we anticipate that torsional buckling occurs when this quantity vanishes, i.e., when $\epsilon=\epsilon_c$ where
\beq
\epsilon_c = - \frac{2 h^2}{a^2\left(1+\nu\right)} + \ed - \edd \quad\textrm{(onset of torsional buckling).} \label{eqn:defnepsilonc}
\eeq
In the right-hand side, the term $-2h^2/[a^2(1+\nu)]<0$ captures the stabilizing effect of the plate's bending modulus (the larger $h^2$, the more negative the critical strain $\epsilon_c$), the term $\ed<0$ captures the stabilizing effect of the longitudinal force (overall tension), and the term $-\edd$ captures the effect of the pre-stress \emph{inhomogeneity} on the twisting rigidity, as can be seen from its definition in Eq.~\eqref{eqn:epsilondags}. The pre-stress is more compressive on the sides of the ribbon (larger $T^2$) than near the centerline (smaller $T^2$), causing $\edd$ to be negative; hence the term $-\edd>0$ in Eq.~\eqref{eqn:ribbonenergydim} points to a \emph{destabilizing} effect of the pre-stress inhomogeneity on torsional buckling.

\subsection{Non-dimensionalization}

Before proceeding to solve the 1D model, we non-dimensionalize by setting
\beqa
&& \cW = Y h a\eta^4 W, \quad \epsilon = \eta^2 E, \quad \epsilon_c = \eta^2 E_c, \quad p = \eta^2 P, \quad \delta = -\ell \eta^2\bar{E}, \nonumber \\
&& \qquad k = \frac{\eta}{a}K, \quad \phi(\ell) = \frac{\ell \eta}{a}\bar{K}, \quad f = Y h a \eta^2 F, \quad m = Y h a^2 \eta^3 M, \label{eqn:nondim}
\eeqa
where we have introduced the slenderness parameter
\beq
\eta \equiv \frac{h}{a\sqrt{12\left(1-\nu^2\right)}}. \label{eqn:defneta}
\eeq
We note that, using the scaling behavior $\eta = O(h/a)$, the scales used to non-dimensionalize $\epsilon$ and $k$ are precisely those reported earlier in Eq.~\eqref{eqn:scalingforms}. We also emphasize the minus sign used to non-dimensionalize the end-shortening $\delta$, in agreement with the usual convention that a contractile strain is counted negative: both $E$ and $\bar{E}$ are negative for compression (when $\delta > 0$), and become more negative as the end-shortening increases.

Using the definition of $\epsilon_c$ in Eq.~\eqref{eqn:defnepsilonc}, and applying the above re-scalings, the strain energy in Eq.~\eqref{eqn:ribbonenergydim} becomes
\beq
W(E,K) = \frac{1}{2}\left[\left(E + \chi P\right)^2 - \chi^2 P^2 + \frac{1}{12}\left(E - E_c\right)K^2 + \frac{1}{320}K^4\right], \label{eqn:ribbonenergy}
\eeq
where $E$, $P$, $E_c$ and $K$ are the re-scaled axial strain, pre-strain, critical strain for torsional buckling, and twisting strain, respectively, and $\chi \in(0,1)$ is the area fraction occupied by the inner (pre-stressed) region. After substituting the expressions in Eq.~\eqref{eqn:epsilondags} for $\ed$ and $\edd$, the critical strain in Eq.~\eqref{eqn:defnepsilonc} becomes, in re-scaled form,
\beq
E_c = -\chi^3 P - 24(1-\nu). \label{eqn:defnEc}
\eeq
The constraints in Eq.~\eqref{eqn:avgekdim} are equivalent to
\beq
\avglength{E} = \bar{E}, \quad \avglength{K} = \bar{K}, \label{eqn:avgEK}
\eeq
i.e., the average values of $E$ and $K$ along the ribbon must match the target values $\bar{E}$ and $\bar{K}$ imposed by the clamps.

\subsection{Convexification and constitutive laws of the 1D model}
\label{sec:ribbonequibmsolns}

\begin{figure}
\centering
\includegraphics[width=\textwidth]{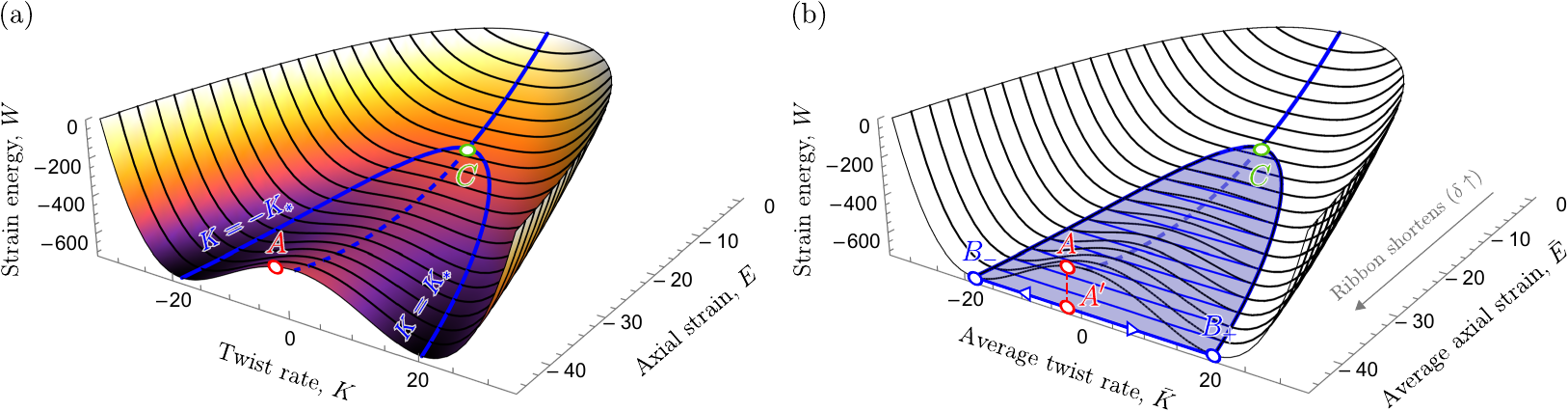} 
\caption{Dimensionless energy landscape according to the ribbon model (here $P = 100$, $\chi = 1/3$, $\nu = 0.49$). (a) Surface plot of the strain energy, $W = \cW/(Y h a\eta^4)$, as a function of the local axial strain, $E = \epsilon/\eta^2$, and twist rate, $K = a k/\eta$, as predicted by Eq.~\eqref{eqn:ribbonenergy}. The point $A$ (red circle) represents a generic point on the surface in the non-convex region; this region is bounded by the curves $K = \pm K_{*}(E)$ (solid blue curves) that emerge from the point $C$ (green circle). (b) Convexified strain energy in terms of the \emph{average} strains $\bar{E} = -\delta/(\ell\eta^2)$ and $\bar{K} = a\phi(\ell)/(\ell\eta)$ that are applied via end-shortening and end-rotation, respectively. The non-convex region $|K| \leq K_{*}$ is replaced by the ruled surface (shaded blue) formed by line segments that join the points at $K=\pm K_{*}$ ($B_{\pm}$, blue circles) for each axial strain. Point $A'$ is the projection of point $A$ onto this surface.}
\label{fig:energylandscape}
\end{figure}

The dimensionless energy landscape as a function of the strain variables $E$ and $K$ is shown in Fig.~\ref{fig:energylandscape}a. This landscape is symmetric as $K \to - K$, as required by Eq.~\eqref{eqn:ribbonenergy}. We observe two valleys (solid blue curves) where ${\partial W/\partial K =0}$, which emerge from the point on the surface where $(E,K)=(E_c,0)$ (labeled $C$). Using Eq.~\eqref{eqn:ribbonenergy}, these valleys are given by
\beq
K = \pm K_{*}(E), \quad K_{*}(E) \equiv \sqrt{\frac{40}{3}\left(E_c-E\right)}, \label{eqn:defnKast}
\eeq
Hence, as the axial strain $E$ quasi-statically decreases from zero (i.e., as the end-shortening increases), equilibrium solutions with non-zero twist are first observed at $E = E_c$ (where $E_c < 0$ from Eq.~\eqref{eqn:defnEc}). Furthermore, as suggested by Fig.~\ref{fig:energylandscape}a, the planar, untwisted solution, $K = 0$, is a local energy minimum as $K$ varies (for fixed $E$) when $E > E_c$ (solid blue curve above point $C$), and a local energy maximum when $E < E_c$ (dashed blue curve). The point $E = E_c$ corresponds to a supercritical pitchfork bifurcation, in which the planar solution becomes unstable to the pair of stable, buckled solutions with opposite chiralities, $K = \pm K_{*}$.

% When the system is in equilibrium, we may identify the partial derivatives of $W$ along $E$ and $K$ with, respectively, the constant (dimensionless) axial force, $F$, and moment, $M$, in the ribbon:
% \beqa
% F(E,K) & = & \pd{W}{E} = E + \chi P + \frac{1}{24}K^2, \label{eqn:force} \\
% M(E,K) & = & \pd{W}{K} = \frac{E + E_c}{12}K + \frac{1}{160}K^3. \label{eqn:moment}
% \eeqa
% Due to the symmetry as $K \to -K$ in the energy \eqref{eqn:ribbonenergy}, we note that buckled solutions must only exist in pairs that have equal and opposite twist rate. Because the axial force $F$ only depends on the twist rate via the $K^2$ term in \eqref{eqn:force}, even if buckled phases co-exist along the ribbon, the associated axial strain $E$ must be the same in both phases. Ignoring gradient effects (and hence boundaries between buckled phases), we deduce that $E$ must be independent of $S$, so that it is everywhere equal to its average:
% \beqn
% E(S) = \bar{E}.
% \eeqn

Because we control only the \emph{average} axial strain $\bar{E}$ and twist rate $\bar{K}$ via the end-shortening and end-rotation, we must account for possible co-existence of buckled phases along the ribbon length. This is achieved by a convexification of the energy landscape in Fig.~\ref{fig:energylandscape}a, the result of which is shown in Fig.~\ref{fig:energylandscape}b. The non-convex part of energy surface, bounded by the curves $K = \pm K_{*}$, is replaced by a ruled surface (shaded blue). This ruled surface is swept by the line segments that join points $K = -K_{*}$ and $K = +K_{*}$ for each $E$.

Physically, the energy convexification may be interpreted as follows. For specified $\bar{E}$ and $\bar{K}$, one possible solution is the homogeneous buckled configuration with $E(S) = \bar{E}$ and $K(S) = \bar{K}$ everywhere along the ribbon. Now suppose that the point $(\bar{E},\bar{K})$ lies inside the non-convex region, as represented by the point $A$ in Fig.~\ref{fig:energylandscape}a. After convexification (Fig.~\ref{fig:energylandscape}b), the point $A$ is projected onto the point $A'$ lying on the ruled surface: the homogeneous solution $A$ is not the global energy minimum, and the system adopts a lower energy state $A'$ by forming a mixture of buckled phases while conserving the average twist rate\footnote{Here we consider only global energy minima. We ignore \emph{metastable} homogeneous solutions (with $E(S) = \bar{E}$ and $K(S) = \bar{K}$ everywhere along the ribbon), which exist in the region bounded by the coexistence curves $K = \pm K_{*}$ and the spinodal curves that emerge from the point $C$ (defined as the locus of inflection points where $\partial^2 W/\partial K^2 = 0$) \citep{jones2002}; using Eq.~\eqref{eqn:ribbonenergy}, this region is $K_{*}/\sqrt{3} < |K| < K_{*}$. Because we always start with zero average twist ($\bar{K} = 0$) before quasi-statically increasing/decreasing $\bar{K}$ in simulations and experiments, these metastable states are not observed.}. Because generators of the ruled surface are evidently parallel to the $K$-axis, the buckled phases in the mixture (corresponding to the points $B_{\pm}$) everywhere have
\beq 
E(S) = \bar{E}, \quad K(S) = \pm K_{*}(\bar{E}). \label{eqn:convexificationrule}
\eeq
The reduced ribbon energy \eqref{eqn:ribbonenergy} therefore predicts the existence of buckled phases that possess opposite chiralities, but, without higher-order gradient terms, cannot describe perversions (phase boundaries). However, if the point $(\bar{E},\bar{K})$ instead lies outside the non-convex region, this conclusion does not hold: since $|\bar{K}| > K_{*}$, it is not possible to satisfy $K = \pm K_{*}$ point-wise while conserving the average value $\bar{K}$ (as required by Eq.~\eqref{eqn:avgEK}). Thus, in this latter case, phase separation does not occur and the system instead adopts the homogeneous phase $(E,K) = (\bar{E},\bar{K})$. 
% In particular, for fixed $E < E_c$, metastable solutions do not exist in a neighborhood of $K = 0$; they also only connect to the branch of inhomogeneous (mixed phase) solutions at $|K| = K_{*}$. 

Using Eqs.~\eqref{eqn:ribbonenergy} and \eqref{eqn:convexificationrule}, the convexified strain energy, labeled $\bar{W}$, can be written in terms of the average strains $\bar{E}$ and $\bar{K}$ as 
\beqa
\bar{W}(\bar{E},\bar{K}) & = & \begin{cases} 
 W(\bar{E},K_{*}) \quad & \mathrm{if} \quad \bar{E} \leq E_c \ \ \mathrm{and} \ \ |\bar{K}| \leq K_{*}, \\
 W(\bar{E},\bar{K}) \quad & \mathrm{otherwise},
\end{cases} \nonumber \\
& = & \begin{cases} 
 (2/9)\bar{E}^2 + \left[\chi P + (5/9)E_c\right]\bar{E} -(5/9)E_c^2 \quad & \mathrm{if} \quad \bar{E} \leq E_c \ \ \mathrm{and} \ \ |\bar{K}| \leq K_{*}, \\
 \bar{E}^2/2 + \chi P\bar{E} + \left(\bar{E} - E_c\right)\bar{K}^2/24 + \bar{K}^4/640 \quad & \mathrm{otherwise}.
\end{cases} \label{eqn:convexenergy}
\eeqa
The corresponding effective constitutive laws for the dimensionless equilibrium force, $F$, and moment, $M$, are 
\beq
F(\bar{E},\bar{K}) = \pd{\bar{W}}{\bar{E}} = \begin{cases} 
(4/9)\bar{E} + \chi P+(5/9)E_c \quad & \mathrm{if} \quad \bar{E} \leq E_c \ \ \mathrm{and} \ \ |\bar{K}| \leq K_{*}, \\ 
\bar{E} + \chi P+\bar{K}^2/24 \quad & \mathrm{otherwise}, 
\end{cases} \label{eqn:forceeqmsol}
\eeq
\beq
M(\bar{E},\bar{K}) = \pd{\bar{W}}{\bar{K}} = \begin{cases} 
0 \quad & \mathrm{if} \quad \bar{E} \leq E_c \ \ \mathrm{and} \ \ |\bar{K}| \leq K_{*}, \\
 (\bar{E}-E_c)\bar{K}/12 + \bar{K}^3/160 \quad & \mathrm{otherwise}. 
\end{cases} \label{eqn:momenteqmsol}
\eeq
We refer to these constitutive laws as `effective' because, in being derived from the convexified strain energy, they account for a mixture of buckled phases inside the non-convex region.

\begin{figure}
\centering
\includegraphics[width=\textwidth]{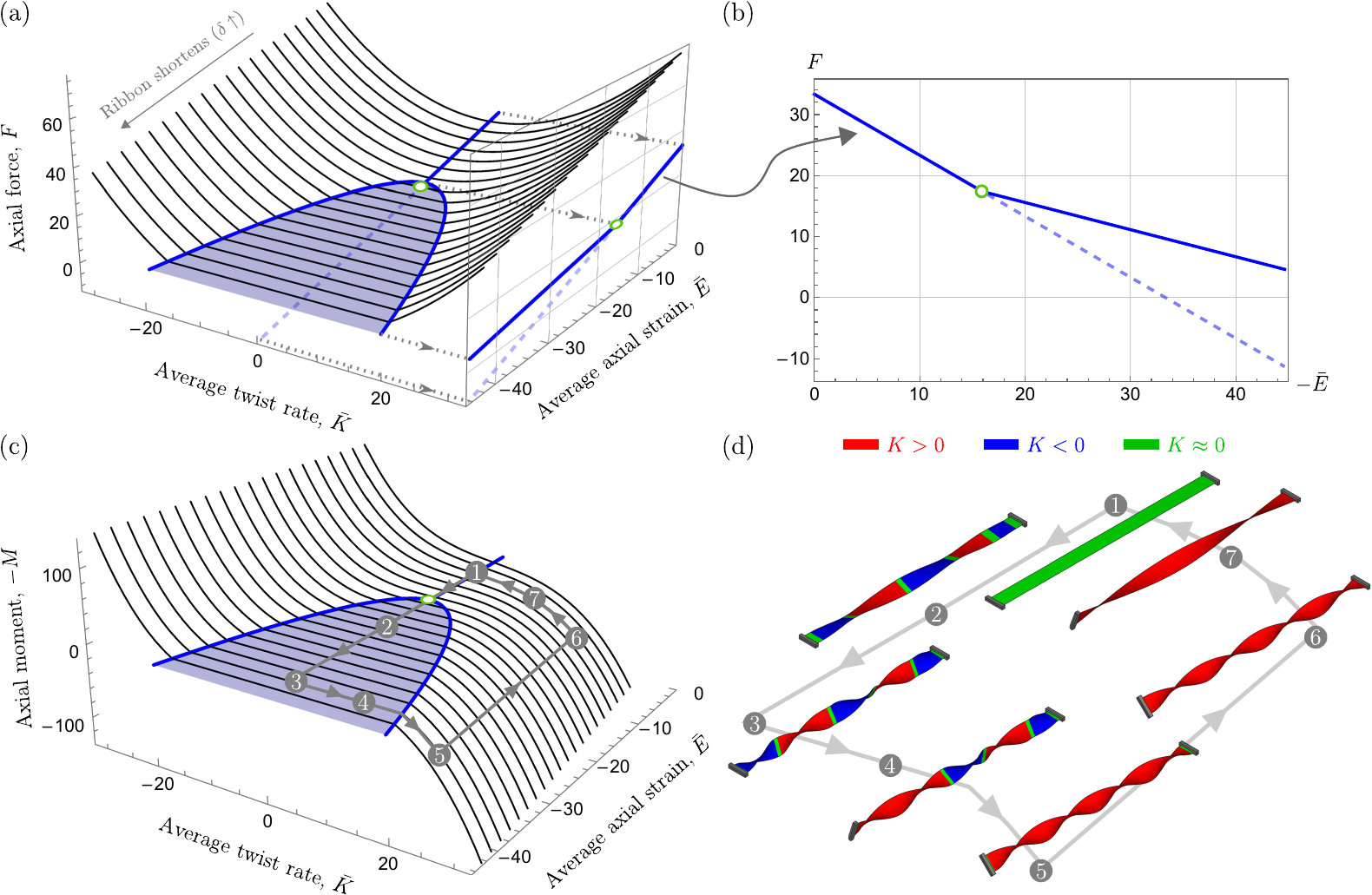}
\caption{Equilibrium behavior according to the convexified ribbon model, i.e.~Eqs.~\eqref{eqn:forceeqmsol}--\eqref{eqn:momenteqmsol} (here $P = 100$, $\chi = 1/3$, $\nu = 0.49$). (a) Surface plot of the dimensionless axial force, $F = f/(Y h a\eta^2)$, as a function of the average axial strain, $\bar{E} = -\delta/(\ell\eta^2)$ and average twist rate, $\bar{K} = a\phi(\ell)/(\ell\eta)$. Each slice (black curves) is an `isotherm' corresponding to constant $\bar{E}$. The convexified region (shaded blue) is bounded by the coexistence curves $\bar{K} = \pm K_{*}$ (solid blue curves); inside this region, a microscopic mixture of buckled phases exist. Taking a vertical slice through $\bar{K} = 0$ yields panel (b). (c) Corresponding surface plot of the negative (dimensionless) axial moment, $-M = -m/(Y h a^2\eta^3)$. (d) Ribbon shapes corresponding to points $1$--$7$ on panel (c), colored according to the sign of the local twist (see legend).}
\label{fig:forcemoment}
\end{figure}

In Fig.~\ref{fig:forcemoment}, we use these expressions to construct surface plots of the force and moment as a function of $\bar{E}$ and $\bar{K}$. More precisely, in panel (c) we plot the \emph{negative} moment, $-M$, as this quantity turns out to be the analog of pressure in the classical theory of first-order phase transitions (see \S\ref{sec:thermoanalogy} below). To help interpret each of these plots, in Fig.~\ref{fig:forcemoment}d we show example shapes of the ribbon corresponding to the points $1$--$7$ on Fig.~\ref{fig:forcemoment}c; these shapes were generated using the FEM simulations discussed in \S\ref{sec:methodsFEM}. (We only draw the shapes for positive applied twist, $\bar{K} > 0$; due to symmetry, the shapes for negative twist are found by simply inverting the chiralities.) These shapes are colored red/blue according to the sign of the local chirality, with green regions corresponding to places where the twist rate is approximately zero. 
%For buckled ribbons, green regions that exist away from the ribbon extremities always occur between neighboring phases of opposite chirality, and hence correspond to perversions.

\subsection{Thermodynamic analogy: torsional buckling as a phase separation process}
\label{sec:thermoanalogy}
The torsional buckling predicted by the ribbon model has many features in common with other critical phenomena described by the classical theory of thermodynamic phase transitions \citep{sears1975,selinger2016}. For example, the mixture of buckled phases observed for $\bar{E} < E_c$ is analogous to the phase separation of a real gas into its liquid and vapor phases when the temperature is lowered below the critical point, as described by van der Waals theory \citep{sears1975}. In fact, Fig.~\ref{fig:forcemoment}c may be directly compared to the standard pressure-volume-temperature (or ``$p$-$V$-$T$'') diagrams for the liquid-gas transition in real substances (ignoring solid phases). Therefore, we have the correspondence:
\beqn
\mathrm{pressure} \longleftrightarrow -M, \quad \mathrm{volume} \longleftrightarrow \bar{K}, \quad \mathrm{temperature} \longleftrightarrow \bar{E}. \label{eqn:pvtcorrespondence}
\eeqn
In particular, the average twist rate $\bar{K}$ (as controlled by end-rotation) changes the relative proportion of the two chiralities, similarly to how the system volume determines the relative proportion of liquid and vapor phases. The level curves along which $\bar{E}$ is constant (black curves in Figs.~\ref{fig:forcemoment}a and \ref{fig:forcemoment}c) can be viewed as isotherms; the bifurcation point $C$ is analogous to the critical point, with $\bar{E} = E_c$ the critical temperature. 

As the system moves quasi-statically between the points $1$--$7$ in Fig.~\ref{fig:forcemoment}c, the ribbon shapes in Fig.~\ref{fig:forcemoment}d also highlight a phenomenon that is well known in the theory of thermodynamic phase transitions: depending on the path taken, it is possible to move between two given points with or without phase separation occurring (for example, compare the paths $1 \to 2 \to 3 \to 4 \to 5$ and $1\to 7\to 6\to 5$).

% For example, consider point $7$ --- corresponding to a `supercritical' buckled state with homogeneous, positive chirality --- and point $5$ --- corresponding to a buckled state also with homogeneous, positive chirality. The system can move between these points either (i) along the path $7\to 6\to 5$, in which case no bifurcation occurs and only the positive chirality phase is observed; or (ii) along the path $7 \to 1 \to 2 \to 3 \to 4 \to 5$, in which case bifurcation occurs at the critical point and both chiralities may temporarily be observed, until the system crosses the boundary of the convexified region between points $4$ and $5$.
 
\section{Methodology: numerical and physical experiments}
\label{sec:methods}

To test the predictions of the extensible ribbon model developed in the previous section, and explore its limits of validity, we also performed computer simulations and desktop-scale experiments. In this section, we begin with a discussion of the numerical simulations in \S\ref{sec:methodsFEM} and then move on to discuss experimental methodology in \S\ref{sec:methodsexpts}. We delay a discussion of the corresponding results until \S\ref{sec:results}.

\subsection{Finite element simulations of pre-stressed ribbons}
\label{sec:methodsFEM}

We conducted simulations based on the finite element method (FEM) using the commercial package Abaqus $6.14$ (Dassault Syst\`{e}mes, Simulia Corp.). The ribbon was meshed using 3D solid elements with quadratic interpolation order (isoparametric, hexahedral elements with reduced integration; type \texttt{C3D20R} in Abaqus with default element controls). We found that using regular, cuboidal elements of side length $h/4$ (i.e., four elements through the thickness) was sufficient to obtain a converged mesh. The values of physical parameters were chosen to exactly match their experimental counterparts (provided below in Eq.~\eqref{eqn:baselinevals}), and were expressed in the millimeter-tonne-second variant of SI units. The discretization comprised $144$ elements in each cross-section and $600$ elements along the longitudinal direction (86400 total elements).

% with uniform side length $0.5\:\mathrm{mm}$

\paragraph{Material constitutive behavior} To compare our simulations to experiments on elastomeric ribbons (described in \S\ref{sec:methodsexpts}), we implemented an isotropic, nearly-incompressible neo-Hookean material model (Poisson ratio $\nu = 0.49$). (Volumetric locking was avoided by using reduced-integration
elements.) While several compressible formulations of the neo-Hookean model have been proposed \citep{pence2015}, we used the default form in Abaqus; this formulation arises as a special case of the more general Mooney-Rivlin solid \citep{bower2009} and has been successfully used to simulate the behavior of elastomers \citep{zhao2019,yan2023}. In terms of the principal stretches $\lambda_i$ ($i = 1,2,3$), the strain energy density (i.e., energy per unit of reference volume) is
\beq
U = \frac{\mu}{2}\left[J^{-2/3}\left(\lambda_1^2 + \lambda_2^2 + \lambda_3^2\right) - 3\right] + \frac{K_b}{2}\left(J-1\right)^2, \label{eqn:strainenergyneohook}
\eeq
where $\mu = Y/[2(1+\nu)]$ and $K_b = Y/[3(1-2\nu)]$ are the shear and bulk modulus, respectively, and $J = \lambda_1\lambda_2\lambda_3$ is the volume ratio.

% We used a nearly-incompressible formulation (Poisson ratio $\nu = 0.49$), to avoid the increased computational cost of hybrid elements that must be used in the fully incompressible case (i.e., $\nu = 0.5$).
%
% \beq
% U = \frac{\mu}{2}\left(\bar{I_1} - 3\right) + \frac{K}{2}\left(J-1\right)^2,
% \eeq
% where $\bar{I_1}$ is the first deviatoric strain invariant (trace) of the right Cauchy-Green deformation tensor
% In terms of the principal stretches $\lambda_i$ ($i = 1,2,3$), we have
% \beqn
% \bar{I_1} = J^{-2/3}\left(\lambda_1^2 + \lambda_2^2 + \lambda_3^2\right), \quad J = \lambda_1\lambda_2\lambda_3.
% \eeqn

\begin{figure}
\centering
\includegraphics[width=\textwidth]{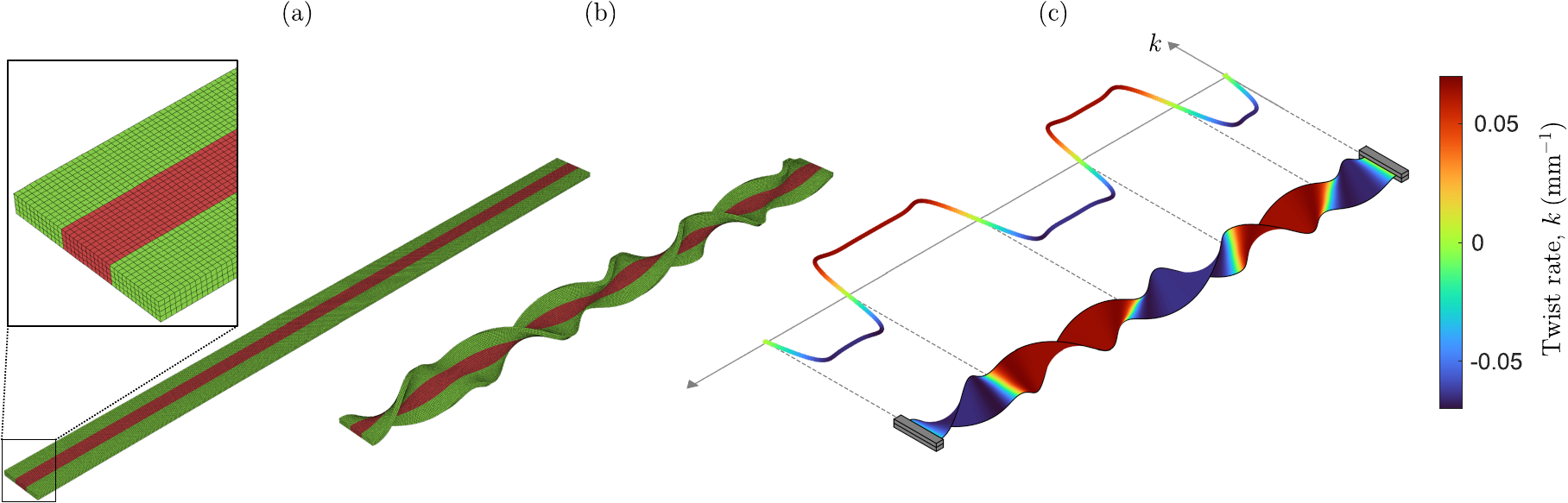} 
\caption{Computational model of elastic ribbons with inhomogeneous, uniaxial pre-stress. The representative example is shown for $\chi = 1/3$, $p = 0.4$, $h = 2\:\mathrm{mm}$, $a = 18\:\mathrm{mm}$, $\ell = 300\:\mathrm{mm}$. (a) 3D view of the ribbon in the fully-unrelaxed configuration. The inset displays a close-up of the regular mesh used in FEM simulations. (b) Corresponding view of a typical shape after torsional buckling occurs ($\delta = 40\:\mathrm{mm}$, $\phi(\ell) = 0$). (c) The local twist rate of the ribbon mid-surface from post-processing the buckled shape shown in panel (b), here plotted as a function of arclength along the deformed centerline.}
\label{fig:methodsFEM}
\end{figure}

\paragraph{Numerical protocols} Instead of explicitly simulating the two-stage stretching and bonding process used to prepare the ribbons (Figs.~\ref{fig:problemdefn}a--d and experimental fabrication in \S\ref{sec:methodsexpts}), we directly constructed the fully-unrelaxed configuration in our simulations; see Fig.~\ref{fig:methodsFEM}a. This step was achieved by first defining the inner and outer regions of the ribbon separately, before merging them so that the mesh naturally retained the internal boundary between the regions. In this way, the pre-stress could be imposed in the inner region as a pre-defined field (via the \texttt{*FIELD} option in Abaqus). For a specified pre-strain $p$, the corresponding uniaxial pre-stress, $\sigma_0$, was derived from the neo-Hookean energy density \eqref{eqn:strainenergyneohook}; for details see \ref{sec:appendixprestress}.

Once the fully-unrelaxed configuration was defined, we conducted FEM analyses under end-shortening and end-rotation. In both loading scenarios, one extremity of the ribbon was clamped by imposing zero displacements at all nodes on the face. At the other extremity, the end-shortening and end-rotation were imposed via a kinematic coupling (Abaqus option \texttt{*COUPLING}) between nodes on the face and a reference point. Throughout, we considered quasi-static loading conditions, using the Abaqus/Standard solver (activating the \texttt{NLgeom} option to incorporate geometric nonlinearities). In addition to the shapes presented earlier in Fig.~\ref{fig:forcemoment}d, Fig.~\ref{fig:methodsFEM}b shows an example buckled shape with the numerical mesh superimposed; here we specified the parameter values in Eq.~\eqref{eqn:baselinevals} with $\chi = 1/3$, $p = 0.4$, $\delta = 40\:\mathrm{mm}$ and $\phi(\ell) = 0$.

To avoid convergence issues at the buckling onset, we seeded the buckling instability using shape imperfections. To obtain the shape imperfections for each pre-strain $p$, we performed a preliminary eigenvalue buckling analysis under incremental changes to the end-shortening, using the planar, unbuckled solution as the base state. The corresponding node displacements for the first ten eigenmodes were then superimposed onto the fully-unrelaxed configuration (\texttt{*IMPERFECTION} option in Abaqus); the modes were scaled to have amplitudes that geometrically decreased with mode number, with the largest amplitude (first buckling mode) having a maximum displacement component equal to $5\%$ of the ribbon thickness. Furthermore, to obtain reliable results that are insensitive to small changes in mesh size and other numerical parameters, it was necessary to include numerical stabilization to deal with the rapid variation in displacements upon buckling (we used adaptive automatic stabilization in Abaqus with default parameter values). By smoothly ramping the rate at which the end-shortening or end-rotation was applied (from zero to the constant value used throughout each loading step), we ensured that the ratio of viscous dissipation energy to the total strain energy always remained less than $5\%$; thus the loading was approximately quasi-static.
  
\paragraph{Post-processing of numerical data} We extracted the raw simulation data output from Abaqus for post-processing in MATLAB, using the toolbox \texttt{Abaqus2Matlab} \citep{papazafeiropoulos2017}. The components of the force and moment resultants were readily obtained from the force and moment exerted on the reference point that was coupled kinematically to one extremity of the ribbon. To obtain the twist rate $k(S)$ for each material coordinate $S \in (0,\ell)$, we first calculated the local director frame using the raw data of node positions along the ribbon mid-surface; this procedure is detailed in \ref{sec:appendixpostprocess}. Figure \ref{fig:methodsFEM}c shows a typical twist profile determined with this approach (plotted as a function of deformed arclength for comparison), corresponding to the buckled shape shown in Fig.~\ref{fig:methodsFEM}b. By determining the director frame, we can also (i) compute the bending strains and, hence, verify that torsional buckling generally occurs with negligible centerline bending, as was assumed in the ribbon model; and (ii) analyze the behavior in regimes where a bending instability occurs (discussed in \S\ref{sec:results}). Once the local twist rate $k(S)$ was determined, we computed its root mean square (RMS), $\sqrt{\avglength{k^2}}$; as in \S\ref{sec:ribbonmodel}, $\avglength{\cdot}$ denotes the average over the ribbon length, $S \in (0,\ell)$.
% While it would have been simpler to compute the twist rate via the angular coordinate of nodes (with respect to a global cylindrical polar coordinate system), our method does not assume that the unit tangent vector is aligned with the $z$-axis. 

% \beq
% \left\langle k^2\right\rangle \equiv \frac{1}{\ell - \delta}\int_0^{\ell-\delta}\left[k(s)\right]^2\id s. \label{eqn:RMSdefn}
% \eeq

\subsection{Experimental methods}
\label{sec:methodsexpts}

We performed precision desktop-scale experiments on ribbons composed of vinyl polysiloxane (VPS, Elite Double 32, Zhermack) --- a silicone-based elastomer whose behavior has been shown to be well approximated, up to strains of around $30\%$, by a neo-Hookean constitutive law with Young's modulus $Y = 1.25\:\mathrm{MPa}$ and Poisson ratio $\nu \approx 0.5$ \citep{baek2021,johanns2021,grandgeorge2021,grandgeorge2022}. We first describe the procedure used to fabricate the ribbons, before discussing the apparatus and methods we employed in the loading tests. 

\begin{figure}
\centering
\includegraphics[width=0.67\textwidth]{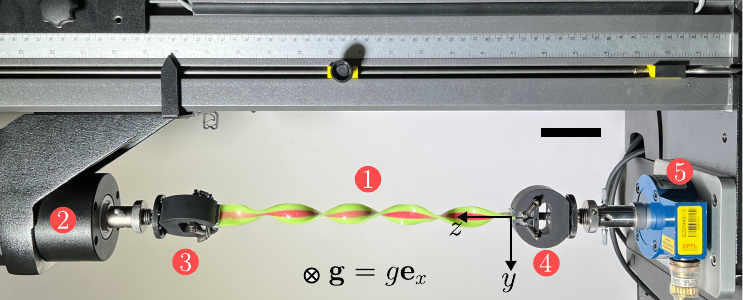} 
\caption{Experimental apparatus used to perform loading tests on pre-stressed elastomeric ribbons (top view; gravity directed into the page). The elastomeric ribbon (1) was attached at one end to a universal testing machine (2) via a movable clamp (3), which applied a precise displacement and rotation about the ribbon axis. The other end was clamped in space (4) and connected to a torsional load cell (5). Scale bar: $5\:\mathrm{cm}$.}
\label{fig:exptsetup}
\end{figure}

\paragraph{Fabrication of elastomeric ribbons} To fabricate VPS ribbons with inhomogeneous, uniaxial pre-stress and specified geometric parameters (given below in Eq.~\eqref{eqn:baselinevals}), we implemented the two-stage fabrication process summarized earlier in Figs.~\ref{fig:problemdefn}a--d. We custom-made molds from acrylic plates (thickness $2\:\mathrm{mm}$; TroGlass Clear Cast, Trotec) that were cut to shape using a laser cutter (Trotec Speedy 400) and bonded together (Acrifix, PLEXIGLAS). For each fabrication stage, a mixture of VPS base and catalyst, in a 1:1 ratio by weight, was placed in a centrifugal mixer (THINKY ARE-250) for a total of $40$ seconds ($20\:\mathrm{s}$ at $2000$ rpm clockwise, $20\:\mathrm{s}$ at 2200 rpm counterclockwise). To minimize imperfections due to air bubbles, the mixture was degassed in a vacuum chamber before being injected into the mold using a syringe \citep{sano2022}. Curing of the polymer mixture took place at room temperature.

The mold for the first fabrication stage featured a uniform rectangular channel with additional end pieces, to cast the inner strip with `clamping blocks' at its ends. The geometry of this channel had to be tailored to each pre-strain $p$ to ensure that the inner strip had the specified thickness $h$, width $\chi a$, and length $\ell$ after uniaxial stretching (specifically, the channel thickness $h_i = (1+p)^{\nu}h$, width $a_i = (1+p)^{\nu}\chi a$ and length $\ell_i=\ell/(1+p)$ with Poisson ratio $\nu = 0.5$, which compensated for the transverse contraction due to Poisson effects). We added a small amount (less than $0.1\%$ by weight) of red silicone pigment (Silc Pig, Smooth-On) to the VPS mixture before curing the inner strip so that the pre-stressed region could be visualized in the finished ribbon. For the second fabrication stage, a second mold held the inner strip under the specified uniaxial strain, using the clamping blocks to place it symmetrically in the center of a wider channel (width $a$) while the outer strips cured around it. This ensured, in turn, that the distribution of pre-stress in the finished ribbons was sufficiently symmetric that they did not develop spontaneous curvature after the mold and clamping blocks were removed. We note that cross-linking of the VPS polymer, at the interface between the inner and outer strips, effectively bonded the strips during the second curing stage without the need for gluing. Each finished sample was left for at least $24$ hours before being used for loading tests.

% To minimize imperfections due to air bubbles, both molds were positioned with their longitudinal axis parallel to the direction of gravity before the de-gassed base/catalyst mixture of VPS was injected into a small hole at their base; the upper-end pieces contained air holes to allow bubbles that had risen to escape.

\paragraph{Apparatus and protocols used for loading tests} Analogously to the FEM simulations, we conducted the two types of mechanical tests (end-shortening and end-rotation; see Figs.~\ref{fig:problemdefn}e--f) on our experimental samples. A photograph of the apparatus is shown in Fig.~\ref{fig:exptsetup}. Each ribbon was initially clamped in its fully-unrelaxed configuration to a universal testing machine (Instron 5943). The sample was clamped with the ribbon's width parallel to the direction of gravity, to minimize sagging due to self-weight. During loading, a torsional load cell (Instron; force capacity $\pm 450\:\mathrm{N}$, torque capacity $\pm 5\:\mathrm{N}\mathrm{m}$) simultaneously measured the axial force and torque. We also recorded the shape of the ribbon using a digital camera (Nikon D850).

\subsection{Parameter values used in the present study}
The values of geometric parameters (defined in \S\ref{sec:problemdefn}) used throughout this paper are
\beq
h = 2\:\mathrm{mm}, \quad a = 18\:\mathrm{mm}, \quad \ell = 300\:\mathrm{mm}, \quad \mathrm{and} \quad
\begin{cases} \:\chi \in \lbrace 1/3, 2/3\rbrace, \quad & p \in [0.04, 0.6] \quad \mathrm{(FEM)}, \\ 
\:\chi = 1/3, \quad & p \in \lbrace 0.2,0.4 \rbrace \quad\:\; \mathrm{(Expts.)}. 
\end{cases} \label{eqn:baselinevals}
\eeq
For each combination of pre-strain $p$ and area fraction $\chi$, we varied the end-shortening in the range $\delta \in [0,40]\:\mathrm{mm}$ (for $\chi = 1/3$) and $\delta \in [0,80]\:\mathrm{mm}$ (for $\chi = 2/3$); larger end-shortenings were required to buckle the ribbon for larger $\chi$. In both simulations and experiments, we used a loading rate of $\dot{\delta} = 0.5\:\mathrm{mm}\:\mathrm{s}^{-1}$ (for $p \geq 0.07$) and $\dot{\delta} = 0.1\:\mathrm{mm}\:\mathrm{s}^{-1}$ (for $p < 0.07$). For the end-rotation tests, we varied the rotation angle in the range $\phi(\ell) \in [0,720]\:\mathrm{deg}$ (simulations) and $\phi(\ell) \in [-720,720]\:\mathrm{deg}$ (experiments) at a constant angular velocity $|\dot{\phi}(\ell)| = 6\:\mathrm{deg}\:\mathrm{s}^{-1}$. In both loading tests, there were no noticeable oscillations of the ribbon, and we verified that changing the rate of loading did not change the results, indicating that the conditions can be regarded as quasi-static. The choice of the above loading rates represents a balance between ensuring quasi-static conditions (in particular, minimizing the viscous dissipation used in the numerical stabilization), while avoiding excessive simulation or experimental times. 

\section{Results of inhomogeneously pre-stressed ribbons under end loads}
\label{sec:results}

In this section, we show that our FEM simulations are able to reproduce the main qualitative features of the torsional buckling exhibited by experimental samples, despite differences in the detailed post-buckled shape. We will also demonstrate that a \emph{quantitative} agreement can be obtained between the results of FEM simulations and experiments when we consider global quantities that are insensitive to the microscopic buckling pattern. By their nature, these quantities do not rely on finite-length (gradient) effects and so can also be predicted by the ribbon model developed in \S\ref{sec:ribbonmodel}, thus enabling a direct comparison between all three types of analysis. We begin by presenting the results for loading by end-shortening in \S\ref{sec:resultsendshort}, and then for end-rotation in \S\ref{sec:resultstwist}.

\subsection{Torsional buckling under end-shortening}
\label{sec:resultsendshort}

\begin{figure}[!ht]
\centering
\includegraphics[width=0.72\textwidth]{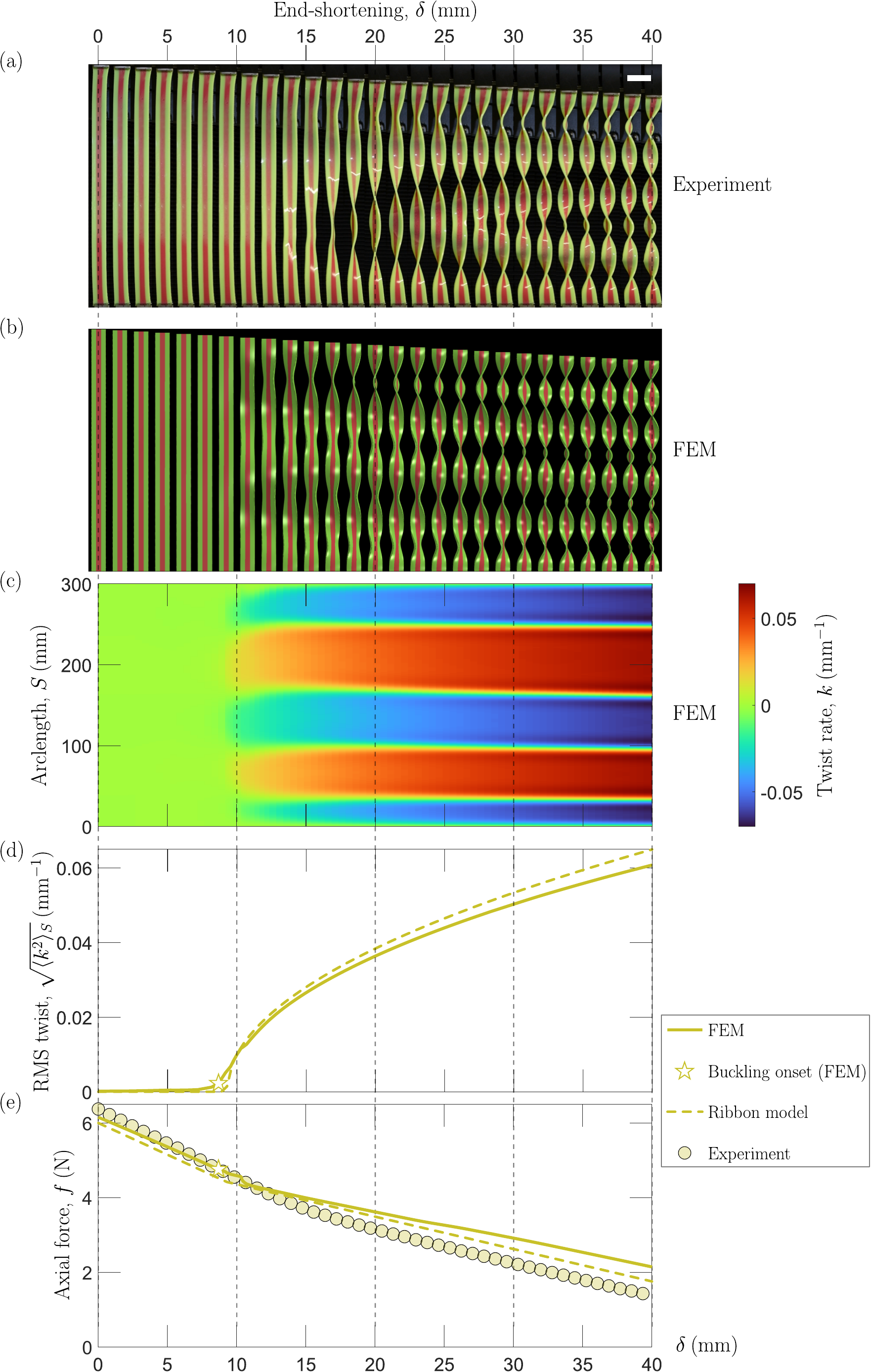} 
\caption{Buckling behavior under pure end-shortening (parameter values in Eq.~\eqref{eqn:baselinevals} with $\chi = 1/3$ and $p = 0.4$). Top panels (a)--(b): Visual comparison of the ribbon shapes obtained (a) experimentally (scale bar: $3\:\mathrm{cm}$); and (b) numerically. In both panels, snapshots are taken at equally-spaced values of the end-shortening in the range $\delta \in [0,40]\:\mathrm{mm}$. (c) Spatial distribution of the twist rate, $k$ (see colorbar), obtained from post-processing the FEM simulation shown in panel (b). (d) Corresponding root mean square (RMS) of the twist rate (solid curve). Also plotted is the equilibrium twist $|k| = (\eta/a)K_{*}$ (dashed curve) predicted by the ribbon model, where $K_{*}$ is calculated using Eq.~\eqref{eqn:defnKast}. (e) Corresponding axial force $f$ (solid curve), together with the model prediction via Eq.~\eqref{eqn:forceeqmsol} (dashed curve) and experimental results (circles).}
\label{fig:endshortsingleprestrain}
\end{figure}

In a first test to validate the FEM simulations against experiments, in Figs.~\ref{fig:endshortsingleprestrain}a--b we compare the sequence of ribbon shapes obtained under end-shortening for a pre-strain $p = 0.4$ and area fraction $\chi = 1/3$. We find excellent qualitative agreement between some features of the simulations and experiments, especially regarding the onset of torsional buckling and subsequent separation into distinct buckled phases. However, we also note differences in the precise buckling pattern that is obtained, including the location and number of perversions: two perversions arise in the experiments as opposed to four in the simulations. We attribute these differences in the buckling pattern to experimental imperfections, as discussed further in \S\ref{sec:ribbonvalidity}.

To quantitatively examine the buckling behavior, we consider the twist rate, $k$, along the ribbon centerline. In Fig.~\ref{fig:endshortsingleprestrain}c, we present a density plot of $k$ as a function of end-shortening $\delta$ and material coordinate (undeformed arclength) $S \in (0,\ell)$ for the same simulation used to generate the shapes in Fig.~\ref{fig:endshortsingleprestrain}b. This $k(\delta,\,S)$ plot was obtained using the post-processing procedure described in \S\ref{sec:methodsFEM} so that each vertical slice through the plot (i.e., a line of constant $\delta$) corresponds to a twist profile akin to that of Fig.~\ref{fig:methodsFEM}c (though now plotted as a function of $S$). Compared with the numerical shapes (Fig.~\ref{fig:endshortsingleprestrain}b), the onset of torsional buckling at $\delta \approx 10\:\mathrm{mm}$ and phase separation is now more clearly visible. As expected, the magnitude of the twist rate associated with each phase grows as the end-shortening increases. For end-shortenings well beyond the onset of buckling, the twist rate is approximately constant throughout each buckled phase, only varying significantly in the narrow perversions that separate neighboring phases and at the clamped extremities of the ribbon. 

% We observe similar features for other values of the pre-strain $p$ and area fraction $\chi$; generally, we find larger $\chi$ is associated with a larger critical end-shortening and a greater number of perversions.

While the density plot, $k(\delta,\,S)$, in Fig.~\ref{fig:endshortsingleprestrain}c helps visualize the twist distribution along the ribbon, it cannot be directly compared to the ribbon model. To enable such a comparison, we consider the root mean square (RMS) of the twist rate, $\sqrt{\avglength{k^2}}$. Because perversions are confined to regions whose length is comparable to the ribbon width $a$ ($ \ll \ell$), the RMS twist rate should provide a good approximation to the magnitude of the preferred twist in the buckled phases (this magnitude is equal in both phases due to the symmetry of the system as $k \to -k$). This is shown in Fig.~\ref{fig:endshortsingleprestrain}d (solid curve) using the simulation data from Fig.~\ref{fig:endshortsingleprestrain}c. We observe excellent agreement with the equilibrium twist predicted by the ribbon model (dashed curve): recalling the non-dimensionalization in Eq.~\eqref{eqn:nondim}, the dimensional prediction is $|k| = (\eta/a)K_{*}$, where $K_{*}$ is defined in Eq.~\eqref{eqn:defnKast}. The sudden growth in the twist rate at buckling closely follows the behavior expected from Eq.~\eqref{eqn:defnKast}, namely $k \propto\sqrt{\delta - \delta_c}$ near the critical end-shortening $\delta_c = -\ell\eta^2 E_c$. While the numerical curve varies smoothly from the planar, untwisted configuration due to the presence of shape imperfections in our FEM implementation (discussed in \S\ref{sec:methodsFEM}), it is helpful to define an empirical buckling onset for FEM simulations --- hereon, we consider the first value of the end-shortening at which the \emph{dimensionless} RMS twist rate, $\sqrt{\avglength{K^2}}$, exceeds $1$, yielding the points represented by the star symbols in Figs.~\ref{fig:endshortsingleprestrain}d--e.

Furthermore, the buckling behavior can be quantified by the axial force in the ribbon. In Fig.~\ref{fig:endshortsingleprestrain}e, we plot the axial force $f$ for the same simulation used for Figs.~\ref{fig:endshortsingleprestrain}b--d (solid curve). We have also superimposed experimental results (circles) and the model prediction $f = Y h a \eta^2 F$ (dashed curve), where $F$ was given by Eq.~\eqref{eqn:forceeqmsol} (setting $\bar{K} = 0$ in the absence of end-rotation). We obtain good agreement between all three sets of data. The change in slope upon torsional buckling, as predicted from the ribbon model (Eq.~\eqref{eqn:forceeqmsol} and Fig.~\ref{fig:forcemoment}b), is clearly evident in the numerical and experimental curves.
% \footnote{The buckling behavior can also be quantified by the axial moment. As predicted by Eq.~\eqref{eqn:momenteqmsol}, in the case of pure end-shortening (zero end-rotation), we find that the axial moment remains close to zero in both simulations and experiments, with small variations due to the appearance of perversions and other finite-length effects. Because these plots are not particularly informative, we discuss only the axial force here.}
% ; these match up well with both the buckling onset determined from the RMS twist plots (star) and the theoretical prediction at $\delta_c $.

\begin{figure}[!t]
\centering
\includegraphics[width=0.93\textwidth]{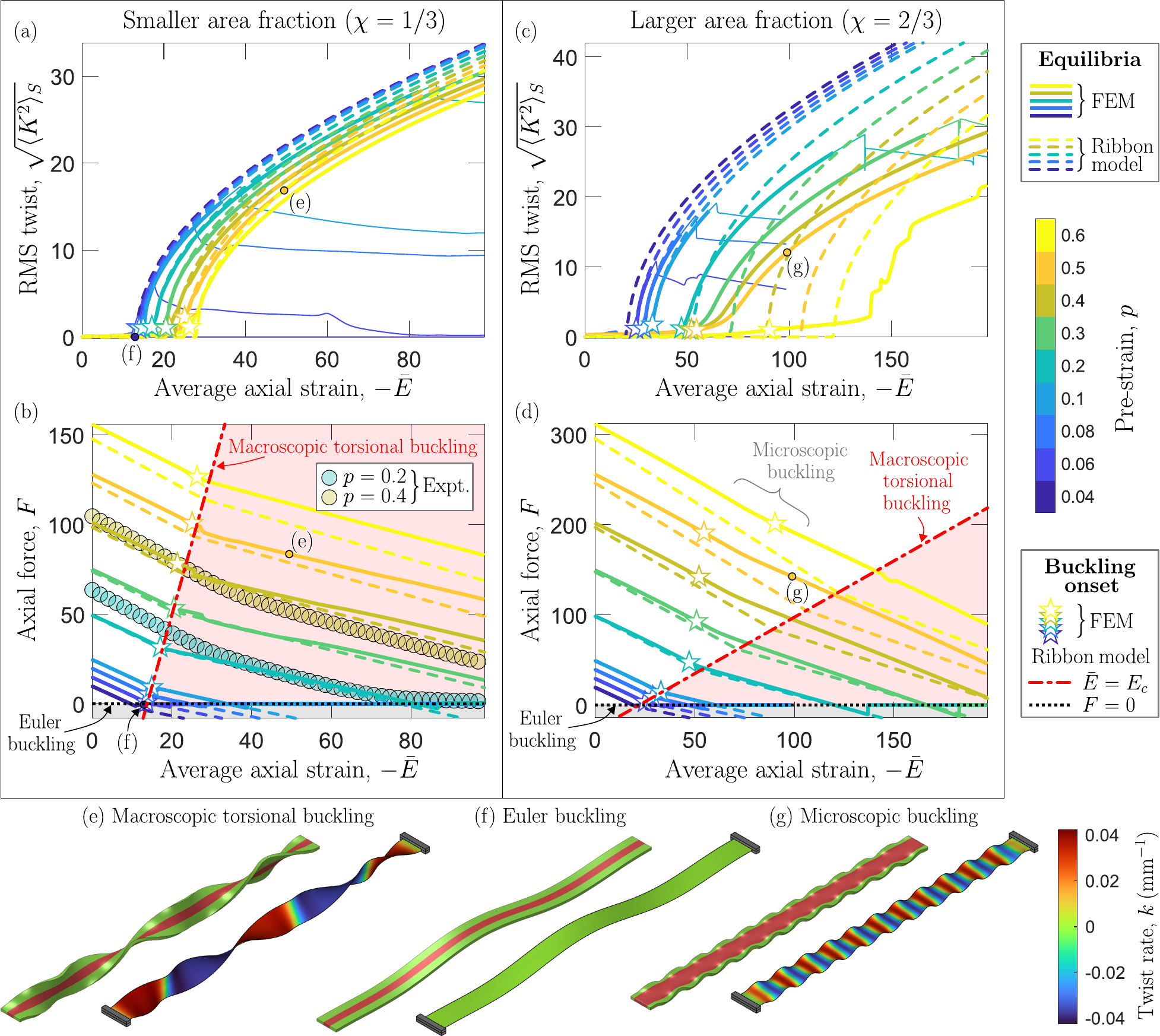}
\caption{Top panels (a)--(d): Behavior under end-shortening for several pre-strains $p \in [0.04,0.6]$ (see colorbar). In all panels we show results from FEM simulations (solid curves), using a reduced line thickness after the point where the axial force crosses zero; the onset of torsional buckling is also highlighted (stars). In addition, we show the corresponding predictions of the ribbon model from Eqs.~\eqref{eqn:defnKast} and \eqref{eqn:forceeqmsol} (dashed curves). (a): Root mean square (RMS) of the dimensionless twist rate, $K = a k/\eta$, as a function of the average axial strain during end-shortening, $\bar{E} = -\delta/(\ell\eta^2)$, using a smaller area fraction, $\chi = 1/3$. (b) Corresponding behavior of the dimensionless axial force, $F = f/(Y h a\eta^2)$, for $\chi = 1/3$. Also plotted is the line $F = 0$ (black dotted line) and the locus of critical points $(\bar{E},F) = (E_c,E_c+\chi P)$ (at which the planar shape bifurcates according to Eq.~\eqref{eqn:forceeqmsol}) as $P$ varies (red dashed-dotted curve); these curves bound the shaded regions where, respectively, Euler-buckled and torsional-buckled shapes are expected. Furthermore, experimental results (circles) are shown for $p = 0.2$ and $p = 0.4$. (c)--(d): As in panels (a)--(b) though with a larger area fraction, $\chi = 2/3$. Bottom panels (e)--(g): Post-buckled ribbon shapes obtained numerically, corresponding to the points labeled (e)--(g) in panels (a)--(d) (pre-strains $p = 0.04$ and $p = 0.5$).}
\label{fig:endshortvaryprestrain}
\end{figure}

\paragraph{Other values of the pre-strain $p$}
In Figs.~\ref{fig:endshortvaryprestrain}a--d, we show the RMS twist rate and dimensionless axial force, $F$, during end-shortening for a range of pre-strains $p \in [0.04,0.6]$ and area fractions $\chi \in \lbrace 1/3,2/3\rbrace$. These plots are presented in dimensionless form, using the non-dimensionalization \eqref{eqn:nondim} introduced in \S\ref{sec:ribbonmodel}. 

Results for a smaller area fraction ($\chi = 1/3$) are shown in Figs.~\ref{fig:endshortvaryprestrain}a--b.

\begin{itemize}

\item In terms of the RMS twist rate (Fig.~\ref{fig:endshortvaryprestrain}a), we generally obtain excellent agreement between the results from FEM simulations (solid curves) and the ribbon model (dashed curves), despite the model being formally valid only for small strains, $p \ll 1$. For clarity, however, here (and in Fig.~\ref{fig:endshortvaryprestrain}c) we have reduced the line thickness of each numerical curve once the axial force becomes negative, indicating that the ribbon is under net compression: soon after this point, the ribbon centerline bends to form an arch shape, similar to classical Euler buckling of a compressed column. The numerical curve then abruptly deviates from the theoretical prediction since the model assumes zero centerline bending.

\item The force-displacement curves from FEM simulations (Fig.~\ref{fig:endshortvaryprestrain}b) generally agree well with the theoretical prediction in Eq.~\eqref{eqn:forceeqmsol}, though we observe a systematic discrepancy for larger pre-strain $p$. Because this discrepancy becomes more significant as $p$ increases, it is likely due to the nonlinear elastic response in the neo-Hookean material model \eqref{eqn:strainenergyneohook} used in FEM simulations (the ribbon model assumes a linear constitutive relation). For each pre-strain $p\geq 0.06$, the numerical curve displays a clear change in slope at the onset of torsional buckling (stars), which matches up well with the theoretical prediction $(\bar{E},F) = (E_c,E_c + \chi P)$ (given by the red dashed-dotted curve as $P=p/\eta^2$ varies). A typical shape of the ribbon after torsional buckling (corresponding to the point labeled (e)) is shown in Fig.~\ref{fig:endshortvaryprestrain}e. However, for the smallest pre-strain, $p = 0.04$, the numerical curve crosses $F = 0$ (black dotted line) before any significant change in slope is observed; the ribbon shape in this case confirms that a bending instability occurs before torsional buckling (see Fig.~\ref{fig:endshortvaryprestrain}f).

\item In addition, Fig.~\ref{fig:endshortvaryprestrain}b displays experimental data for the pre-strains $p = 0.2$ and $0.4$ (circles). These generally agree well with the analytical and numerical results except when the axial force becomes very small, when we observed significant sagging of the ribbon due to gravitational forces in experiments. 

\end{itemize}

Results for a larger area fraction ($\chi = 2/3$) are shown in Figs.~\ref{fig:endshortvaryprestrain}c--d.

\begin{itemize}

\item When the pre-strain $p \lesssim 0.2$, we again observe good agreement between theory and FEM simulations in terms of the RMS twist rate (Fig.~\ref{fig:endshortvaryprestrain}c) and axial force (Fig.~\ref{fig:endshortvaryprestrain}d). In particular, the empirically-determined onset of torsional buckling (stars) is well approximated by $\bar{E} = E_c$ (defined in Eq.~\eqref{eqn:defnEc}) for each of these pre-strains (red dashed-dotted curve in Fig.~\ref{fig:endshortvaryprestrain}d). Similar to the $\chi = 1/3$ case, as the end-shortening is increased further, the numerical curves suddenly deviate from the theory soon after the point where the axial force crosses zero (indicated by a reduced line thickness) and a bending instability occurs (Euler buckling).

\item 
However, a different picture is obtained for larger pre-strains $p \gtrsim 0.2$: along these numerical curves, torsional buckling sets in much earlier than what is predicted by the ribbon model (Fig.~\ref{fig:endshortvaryprestrain}d), leading to significant discrepancies throughout the entire range of end-shortening. By comparing a typical ribbon shape in this regime (Fig.~\ref{fig:endshortvaryprestrain}g with the corresponding shape for $\chi = 1/3$ (Fig.~\ref{fig:endshortvaryprestrain}e), we find that the buckling mode is qualitatively different, being \emph{microscopic} in nature for $\chi = 2/3$: the wavelength is comparable to the ribbon width, $a$, leading to a large number of perversions (typically $15$ or more). Furthermore, for even larger $p$ (not shown here), the buckling shifts to being microscopic and of \emph{bending} type, resembling a wrinkling instability. As discussed in \S\ref{sec:ribbonvalidity}, both types of microscopic buckling lie outside the validity limit of the ribbon model.

\end{itemize}

In summary, we have a competition between macroscopic helical buckling and Euler buckling for a smaller area fraction ($\chi=1/3$), and a competition between these two modes plus microscopic buckling for larger area fraction ($\chi=2/3$).

\subsection{Behavior of pre- and post-buckled ribbons under end-rotation}
\label{sec:resultstwist}

%While applying a pure end-shortening (with zero end-rotation) is useful to understand the onset of torsional buckling and subsequent phase separation, 

\begin{figure}[!t]
\centering
\includegraphics[width=0.75\textwidth]{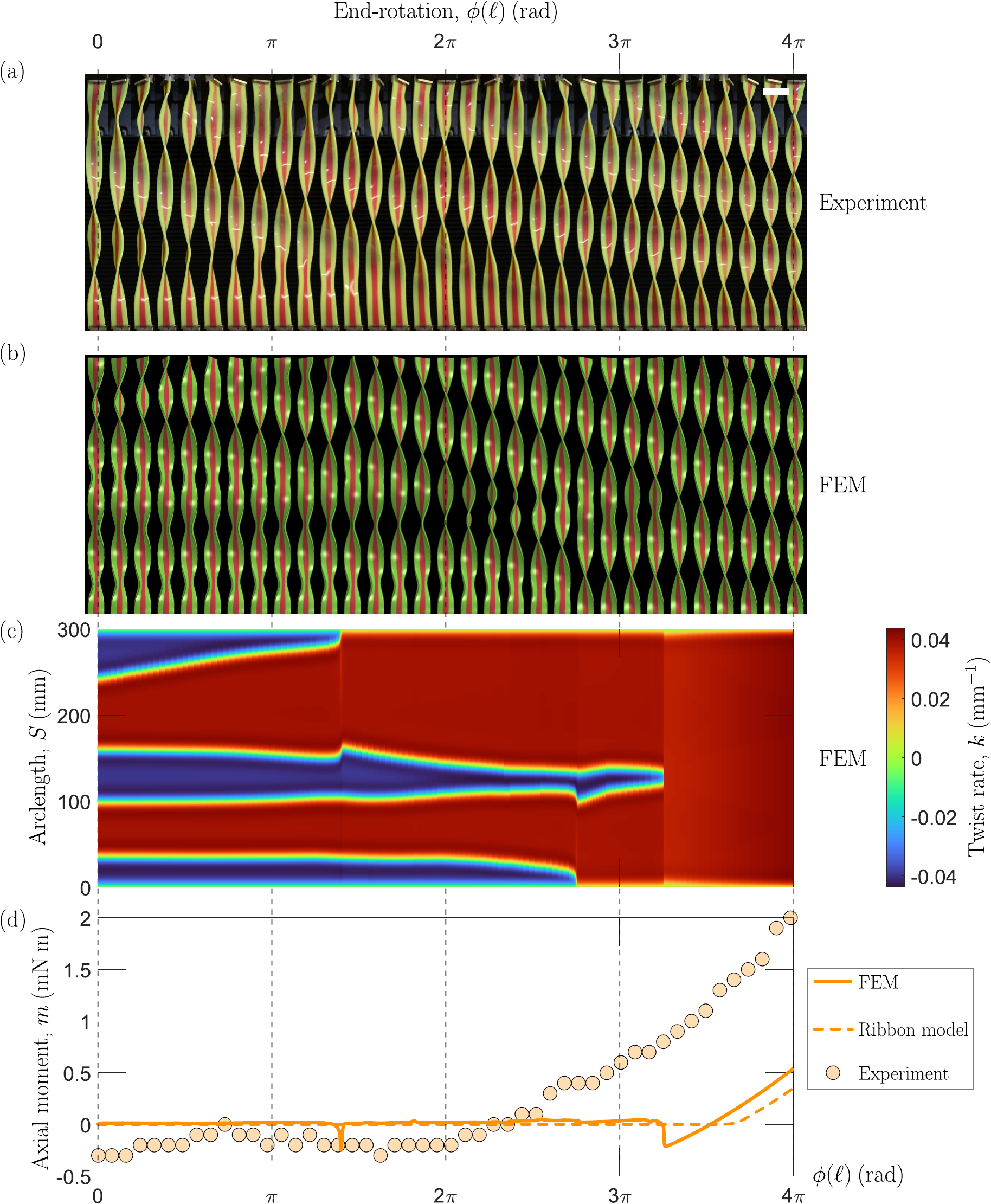} 
\caption{Behavior of the end-shortened ribbon under subsequent end-rotation ($\chi = 1/3$, $p = 0.4$, $\delta = 20\:\mathrm{mm}$). Top panels (a)--(b): Visual comparison of the ribbon shapes obtained (a) experimentally (scale bar: $3\:\mathrm{cm}$); and (b) numerically. In both panels, snapshots are taken at equally-spaced values of the end-rotation up to two complete revolutions, i.e.~$\phi(\ell)\in[0,4\pi]\:\mathrm{rad}$. (c) Spatial distribution of the twist rate, $k$ (see colorbar), obtained from post-processing the FEM simulation shown in panel (b). (d) Corresponding behavior of the axial moment, $m$ (solid curve). Also plotted is the prediction of the ribbon model, computed using Eq.~\eqref{eqn:momenteqmsol} (dashed curve), and experimental results (circles).}
\label{fig:rotationsingleprestrain}
\end{figure}

Next, we consider how the relative proportion of the buckled phases depends on the average applied twist. As in the case of end-shortening, we make a first comparison between experiments and FEM simulations during end-rotation by analyzing the sequence of ribbon shapes; see Figs.~\ref{fig:rotationsingleprestrain}a--b. As in Fig.~\ref{fig:endshortsingleprestrain}, here we have taken $p = 0.4$ and $\chi = 1/3$, though now we fix the end-shortening at $\delta = 20\:\mathrm{mm}$. We obtain a similar picture: the simulations capture the key qualitative features of the experimental system despite differences in the detailed buckling pattern. As the end-rotation increases, we observe the propagation and annihilation of perversions in both the experiment and simulation, with the system eventually reaching a homogeneous state characterized by a single chirality.

For the numerical simulation shown in Fig.~\ref{fig:rotationsingleprestrain}b, the density plot of the twist rate $k(\phi(\ell),\,S)$ (with end-rotation $\phi(\ell)$ now being the second variable, in addition to the material coordinate $S$) is shown in Fig.~\ref{fig:rotationsingleprestrain}c. This plot shows how the four perversions that are initially present propagate and abruptly disappear as they collide with the clamped boundaries and each other. Note that remote parts of the ribbon, including other perversions, undergo sudden re-arrangements after such events. After the system reaches a single, global chirality for large rotations, the associated twist rate becomes uniform along the ribbon (except near the extremities due to boundary effects). 

% We have verified that the mean value of the twist rate equals that imposed by the end-rotation to within numerical errors. 

In Fig.~\ref{fig:rotationsingleprestrain}d, we plot the corresponding FEM-computed axial moment, $m$ (solid curve), as a function of end-rotation, together with experimental results (circles) and the prediction  $m = Y h a^2 \eta^3 M$ from the ribbon model (dashed curve), where $M$ is evaluated using Eq.~\eqref{eqn:momenteqmsol}. Here, we obtain larger quantitative differences compared to the analogous load-displacement curve during end-shortening (Fig.~\ref{fig:endshortsingleprestrain}e). Nevertheless, the three data sets exhibit the same qualitative features, including the plateau where $m\approx 0$ for small rotations and subsequent stiffening for $\phi(\ell) \gtrsim 3\pi\:\mathrm{rad}$.

\begin{figure}
\centering
\includegraphics[width=\textwidth]{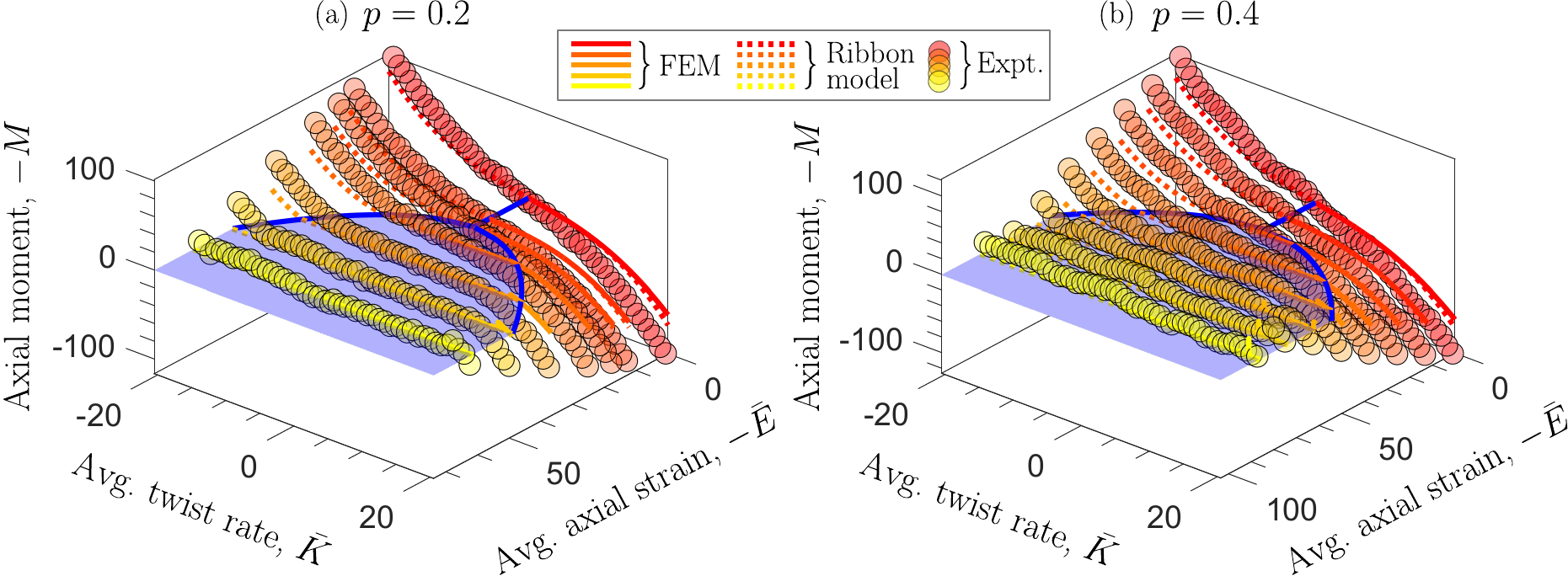} 
\caption{The negative (dimensionless) axial moment, $-M = -m/(Y h a^2\eta^3)$, during end-rotation for several values of the end-shortening (indicated by different colors) and pre-strain (a) $p = 0.2$; and (b) $p = 0.4$ (area fraction $\chi = 1/3$). In each panel, we show results from FEM simulations (solid curves; $\bar{K} > 0$ only), experiments (circles) and predictions of the ribbon model using Eq.~\eqref{eqn:momenteqmsol} (dotted curves). The coexistence curves $\bar{K} = \pm K_{*}$ are superimposed (blue curves), which bound the convexified region (shaded blue) where a mixture of buckled phases is predicted.}
\label{fig:rotationvaryprestrain}
\end{figure}

\paragraph{Other values of the pre-strains $p$}
The negative axial moment during end-rotation for several end-shortenings is shown in Fig.~\ref{fig:rotationvaryprestrain} (plotting dimensionless quantities as defined in Eq.~\eqref{eqn:nondim}), for an area fraction $\chi = 1/3$ and pre-strain $p = 0.2$ (Fig.~\ref{fig:rotationvaryprestrain}a) and $p = 0.4$ (Fig.~\ref{fig:rotationvaryprestrain}b). These plots are similar to the theoretical surface presented in Fig.~\ref{fig:forcemoment}c. In particular, the predicted plateau in the convexified region (highlighted blue) is evident in the numerical and experimental results.
%; the small jumps in this region correspond to sudden re-arrangement of perversions and other finite-length effects. 
For each end-shortening, the value of $\bar{K}$ at which the moment starts to increase rapidly generally matches up well with the coexistence curves $\bar{K} = \pm K_{*}$ (blue curves). However, for values $\bar{K} \gtrsim 15$, the ribbon model systematically underestimates the numerical data when large twisting strains are encountered; similar to the discrepancy observed in the axial force for larger pre-strains in Figs.~\ref{fig:endshortvaryprestrain}b and \ref{fig:endshortvaryprestrain}d, we expect that the discrepancy observed at large twist is due to nonlinear terms in the neo-Hookean constitutive law that are neglected by the ribbon model. 
% We note that similar trends are observed for the axial force $F$ during end-rotation (not presented here).

\section{Limitations of our ribbon model}
\label{sec:ribbonvalidity}
In comparing the quantitative predictions of the ribbon model with FEM simulations and experiments in \S\ref{sec:results}, we generally found good agreement with macroscopic quantities that are insensitive to the precise buckling pattern --- such as the RMS twist rate and axial force --- except in two regimes. First, a macroscopic instability occurs when the axial force transitions from tensile to compressive, which is characterized by significant centerline bending (Euler buckling). This bending instability is generally encountered as the end-shortening $\delta$ is increased beyond the onset of torsional buckling, but, for sufficiently small pre-strain $p$, centerline bending occurs before torsional buckling (Fig.~\ref{fig:endshortvaryprestrain}f); in the latter case, no twisting instability is observed, even as the end-shortening is increased further. In the second regime, generally encountered for larger $\chi$ and large pre-strain $p$, the initial instability is microscopic (see Fig.~\ref{fig:endshortvaryprestrain}g): the buckling wavelength $\lambda \sim a$ instead of $\lambda\sim \ell$, being associated with a large number of perversions (or wrinkles for even larger $p$, when the microscopic instability shifts to being of bending type) that persist upon further end-shortening.

The two regimes mentioned above cannot be described by our ribbon model due to two assumptions made in \S\ref{sec:ribbonmodel}: (i) centerline bending is negligible, and (ii) variations in the strains occur on lengthscales much larger than the ribbon width. While it would be possible to incorporate centerline bending into the ribbon model (recall the discussion at the start of \S\ref{sec:ribbonmodel}), the assumption (ii) lies at the heart of the dimension reduction and cannot be relaxed. Indeed, as well being unable to predict microscopic instabilities, the ribbon model cannot describe deformations in the vicinity of perversions nor finite-length effects such as the number/location of perversions. In future work, we will try to account for the microscopic instabilities by returning to the plate model used as a starting point for the dimension reduction.

% While this may be achieved using the dimensionless equilibrium solutions presented in \S\ref{sec:ribbonequibmsolns},

\subsection{Competition between torsional and Euler buckling}
Equations \eqref{eq:OnsetOfEulerBuckling} and \eqref{eqn:defnepsilonc} yield the critical strain for Euler versus torsional buckling, respectively. These two values are mutually exclusive as they both characterize the stability of the same flat configuration: we should limit attention to whichever instability occurs first during end-shorting. Specifically, torsional buckling occurs before Euler buckling when $\epsilon_c>\ed$ (recall that both quantities are negative), which corresponds to 
%While a complete treatment of the buckling behavior when assumptions (i) or (ii) do not apply is beyond the scope of this paper, we may still use the ribbon model developed here to determine when assumption (i) breaks down. For the sake of generality, we use the dimensional strain energy, $\cW$, given earlier in Eq.~\eqref{eqn:ribbonenergydim}, as our starting point: this incorporates an arbitrary distribution of (uniaxial) pre-stress via the two strains $\ed$ and $\edd$ defined in \eqref{eqn:epsilondags} (our analysis from \S\ref{sec:ribbonequibmsolns} onwards uses the particular form of $\ed$ and $\edd$ for the pre-stress in \eqref{eqn:prestress}). Recall from Eq.~\eqref{eq:OnsetOfEulerBuckling} that, in the absence of twist ($k = 0$) and in the infinite-length limit ($\ell \gg a$), we expect Euler-buckling to occur when $\epsilon=\ed$ \cite[as occurs for the bi-strip analyzed by][]{lestringant2017}. On the other hand, the critical strain at which the planar state becomes unstable with respect to twisting, labeled $\epsilon_c$, is the one that makes the incremental twisting modulus appearing in the square bracket in \eqref{eqn:ribbonenergydim} zero: it is given by
%\beq
%\epsilon_c = - \frac{2 h^2}{a^2(1+\nu)} + (\ed - \edd).  %\label{eqn:defnepsilonc}
%\eeq
%The competition between Euler buckling and torsional buckling is then governed by the relative values of $\epsilon_c$ and $\ed$. In particular, as the ribbon shortens and $\epsilon$ becomes more negative, torsional buckling occurs \emph{before} a bending instability when $\epsilon_c > \ed$. This is equivalent to
\beq
\edd + \frac{2 h^2}{a^2(1+\nu)} < 0. \label{eqn:generaltwistvsbend}
\eeq
Physically, this states that the destabilizing effect of non-uniform pre-stress (as encapsulated in $\edd$) outweighs the stabilizing effect of bending stiffness.

For the particular distribution of pre-stress considered in this paper, i.e., Eq.~\eqref{eqn:prestress}, we have $\edd = -\chi\left(1-\chi^2\right)p$ and the above condition can be re-arranged to
\beq
p >  p_{c_1} \quad \mathrm{where} \quad p_{c_1} \equiv \frac{2 h^2}{a^2(1+\nu)\chi\left(1-\chi^2\right)}\quad\textrm{(torsional buckling earlier than Euler buckling)}.  \label{eqn:defnpcrit1}
\eeq
In dimensionless terms, it follows from Eq.~\eqref{eqn:forceeqmsol} that this is precisely the statement $F(E_c,0) > 0$, i.e., the axial force is positive (tensile) at the onset of torsional buckling when $\bar{K} = 0$; here the critical value $E_c = \epsilon_c/\eta^2$ is as defined in Eq.~\eqref{eqn:defnEc}.

% Defining the re-scaled pre-strain $P = p/\eta^2$ as in \S\ref{sec:ribbonmodel}, where $\eta$ was defined in Eq.~\eqref{eqn:defneta}, the macroscopic bending-twisting transition occurs at the critical pre-strain
% \beq
% P_{c_1} = \frac{24(1-\nu)}{\chi(1-\chi^2)}.
% \label{eqn:defnPcrit1}
% \eeq

% We performed a linear stability analysis of the planar configuration in the F\"{o}ppl-von-K\'{a}rm\'{a}n plate model. This calculation, while entirely analytical, is somewhat technical and will be presented in a subsequent publication. The key feature is that it predicts both the type of buckling mode (bending or twisting)  encountered at onset as the end-shortening is quasi-statically increased, and whether the mode is macroscopic (defined by a wavelength $\lambda \sim \ell$, the overall system size) or microscopic ($\lambda \sim a$).  The linear stability analysis predicts that the macroscopic-microscopic transition occurs at the pre-strain
% \beq
% P_{c_2} = \frac{-(1+\chi ) \left[25\nu-11+3(1-\nu )\chi^2\right] + \sqrt{(1+\chi)^2\left[25\nu -11+3(1-\nu)\chi^2\right]^2+4(1-\nu)(27+71\nu )\left(11-2\chi-\chi^2\right)}}{\chi ^3(1-\chi)\left(11-2\chi -\chi^2\right)/4}.
% \label{eqn:defnPcrit2}
% \eeq

\subsection{Phase diagram of buckling} 
\begin{figure}[h!]
\centering
\includegraphics[width=\textwidth]{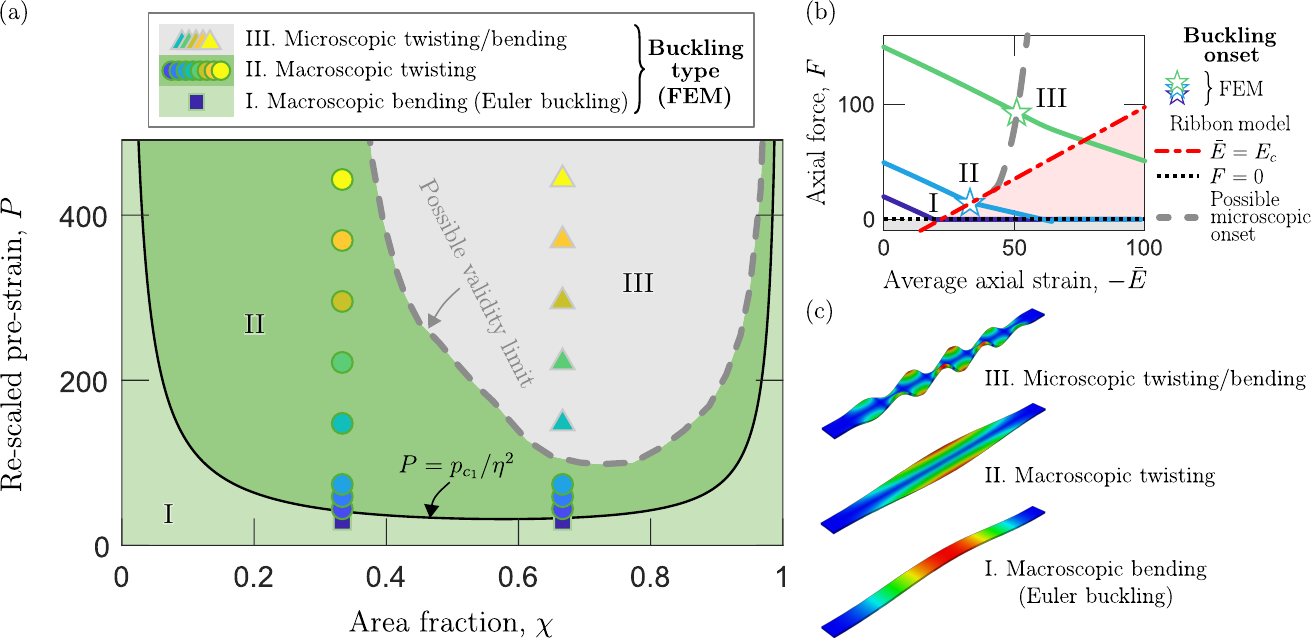} 
\caption{(a) Phase diagram of the instability that is first encountered when the fully-unrelaxed configuration is subject to end-shortening. We plot points $(\chi,P)$ at the parameter values \eqref{eqn:baselinevals} used in FEM simulations ($\chi \in \lbrace 1/3, 2/3\rbrace$ and $P = p/\eta^2$ where $p \in [0.04,0.6]$), where symbols correspond to the type of instability observed (see legend; here the fill color refers to the pre-strain $p$ as in Figs.~\ref{fig:endshortvaryprestrain}a--d). Also plotted is the predicted boundary $P = p_{c_1}/\eta^2$ (black curve) separating regions I and II (from Eq.~\eqref{eqn:defnpcrit1}), and a freehand guess of the boundary between regions II and III (gray dashed curve). (b) Dimensionless axial force, $F = f/(Y h a\eta^2)$, as a function of the average axial strain, $\bar{E} = -\delta/(\ell\eta^2)$, for FEM simulations in each of regions I--III in panel (a) (data from Fig.~\ref{fig:endshortvaryprestrain}d with $\chi = 2/3$ and $p = 0.04$, $0.1$, $0.3$). The red dashed-dotted curve is the locus of critical points $(\bar{E},F) = (E_c,E_c+\chi P)$ as $P$ varies, at which the ribbon model predicts torsional buckling (from Eq.~\eqref{eqn:forceeqmsol}). The gray dashed curve is a guess of the boundary at which microscopic buckling occurs. (c) Representative buckling mode(s) obtained from FEM simulations using an eigenvalue buckling analysis, drawn with arbitrary amplitude.}
\label{fig:ribbonvalidity}
\end{figure}

Combining the above discussion with our numerical results from \S\ref{sec:results}, we can construct a phase diagram of the initial instability that occurs as the end-shortening is quasi-statically increased (starting in the fully-unrelaxed configuration); this is shown on the $(\chi,P)$-plane in Fig.~\ref{fig:ribbonvalidity}a, where $\chi$ is the area fraction occupied by the pre-stressed region and $P = p/\eta^2$ is the re-scaled pre-strain (with $\eta$ as defined in Eq.~\eqref{eqn:defneta}). Here, we have plotted points corresponding to the values of $\chi$ and $P$ used in the FEM simulations (reported in Eq.~\eqref{eqn:baselinevals}), with symbols corresponding to the type of instability observed in \S\ref{sec:resultsendshort}: these are macroscopic bending (Euler-type) in region I (squares), macroscopic twisting-type in region II (circles), and microscopic twisting/bending-type in region III (triangles). In addition, we have included the boundary of regions I and II predicted by the 1D model (black curve), given by $P = p_{c_1}/\eta^2$, using the expression for $p_{c_1}$ in Eq.~\eqref{eqn:defnpcrit1}. The square symbols at the bottom of the each stack of symbols, corresponding to $p=0.04$, both fall below this boundary, in region I. This is consistent with our observation in \S\ref{sec:resultsendshort} of an Euler instability for this value of $p$.

To help interpret the diagram, in Fig.~\ref{fig:ribbonvalidity}b we show typical force-displacement curves, obtained from FEM, for each of the regions I--III (these are reproduced from Fig.~\ref{fig:endshortvaryprestrain}d for $\chi = 2/3$ and various pre-strains $p = 0.04$, $0.1$, $0.3$). Figure \ref{fig:ribbonvalidity}c also displays representative buckling modes. Euler buckling (I) occurs when the force-displacement curve (dark blue curve in the lower left corner of Fig.~\ref{fig:endshortvaryprestrain}d) reaches $F = 0$ (black dotted line) before it intersects the critical strain $\bar{E}=E_c$ for torsional buckling (red dashed-dotted curve). The ribbon model can predict the onset of Euler buckling, Eq.~\eqref{eq:OnsetOfEulerBuckling}, though a description of the post-buckled shape would require centerline bending to be incorporated into the model. In region II, macroscopic torsional buckling takes place at the critical strain $\bar{E}=E_c$ predicted by the ribbon model (intersection of the light blue and red dashed-dotted curves); the model is also valid in the post-buckled regime, provided a secondary bending instability does not occur. As discussed in \S\ref{sec:ribbonvalidity} above, the onset of microscopic buckling (III) (nor the post-buckled behavior) cannot be predicted by the 1D model (green star symbol lying above the red dashed-dotted curve).

%Furthermore, Fig.~\ref{fig:ribbonvalidity}c displays representative buckling modes for each of these cases, obtained from FEM simulations using an eigenvalue buckling analysis (similar to that discussed in \S\ref{sec:methodsFEM} to obtain shape imperfections).

In addition, for illustrative purposes, we have plotted a \emph{speculative} boundary between regions II and III on Fig.~\ref{fig:ribbonvalidity}a, as well as a guess of the locus of critical points at which microscopic buckling occurs on Fig.~\ref{fig:ribbonvalidity}b (gray dashed curves). Together, these form the limit of validity of the ribbon model. We emphasize that these boundaries are freehand guesses. We expect that their precise form may be determined by a linear stability analysis of the F\"{o}ppl-von-K\'{a}rm\'{a}n plate model, which will be considered in a future publication.
    
\section{Discussion and conclusions}
\label{sec:discussionconclusions}

\subsection{Summary of findings}
\label{sec:summary}

In this paper, we have studied elastic ribbons subject to uniaxial pre-stress that is inhomogeneous in the cross-section. In our experiments (discussed in \S\ref{sec:methodsexpts}), we fabricated these ribbons using a stretch-and-bond procedure (\emph{cf.}~Figs.~\ref{fig:problemdefn}a--d) to yield a pre-stress that is non-zero only in a central region of relative area $\chi$. More broadly, a similar distribution of pre-stress could equally arise in an initially unstressed ribbon, in which the inner region contracts relative to the outer region due to active changes such as growth or inflation \citep{koehl2008,siefert2019,gao2020,siefert2020a,moulton2020}. 

Because the pre-stress is mirror-symmetric about the ribbon centerline, classical 1D models predict zero spontaneous curvature and twist and so are unable to describe the torsional instability. Therefore, the ribbon serves as a model system to test and apply novel dimension-reduction tools when standard theories do not apply. In \S\ref{sec:ribbonmodel} we adapted the extensible ribbon model recently developed by \cite{audoly2021b} to incorporate pre-stress. This model successfully captures the salient features of the system observed in experiments and numerical simulations: these include when buckled phases co-exist under a controlled end-to-end displacement and rotation, and the preferred values of the twist rate (away from boundaries and perversions). As well as neglecting strain gradients (and hence microscopic instabilities) and describing only solutions with zero centerline bending, a key assumption of the ribbon model is that the strains are small; in particular, the pre-strain satisfies $p \ll 1$. While the assumption $p \ll 1$ may appear to be overly restrictive, we obtained good quantitative agreement with experiments and FEM simulations for pre-strains up to $p = 0.6$. In addition, the non-dimensionalization introduced in Eq.~\eqref{eqn:nondim} indicates that the system depends on $p$ only via the re-scaled pre-strain $P = p/\eta^2$, where $\eta = O(h/a)$ is the slenderness parameter defined in Eq.~\eqref{eqn:defneta}. It is for this reason that the phase diagram in Fig.~\ref{fig:ribbonvalidity}a is plotted with $P$ on the vertical axis, and the minimum pre-strain for torsional buckling to occur before Euler buckling, $p_{c_1}$, scales as $h^2/a^2$ (recall Eq.~\eqref{eqn:defnpcrit1}). It follows that the torsional instability can be observed with \emph{arbitrarily small} pre-strain $p$ by using a strip with sufficiently small thickness-to-width ratio, $h/a$.

Our ribbon system provides a desktop-scale elastic analog to classical thermodynamic phase separation (e.g., of a real gas into its liquid and vapor phases). Specifically, the mixture of buckled phases occurring for sufficiently negative axial strain (as applied via end-shortening) is analogous to lowering the temperature below the critical point. The average twist rate (applied via end-rotation) changes the relative proportion of the two phases, similarly to how the system volume determines the relative proportion of liquid and vapor phases; the negative axial moment, $-m$, which is the conjugate thermodynamic variable to the twist rate, is then analogous to the pressure. Thus, the moment-twist-strain plots, as predicted by the ribbon model (Fig.~\ref{fig:forcemoment}c) and obtained from numerical and experimental data (Fig.~\ref{fig:rotationvaryprestrain}), are analogous to standard pressure-volume-temperature (``$p$-$V$-$T$'') diagrams \citep{sears1975}. We also note that the convexification of the strain energy (Fig.~\ref{fig:energylandscape}b) is similar to the Maxwell construction (also referred to as the equal-area rule) \citep{clerk1875}. Similar constructions have been used in other problems in elasticity involving co-existing phases in a spatially-extended system, such as bulges in a hyperelastic cylindrical membrane \citep{chater1984}; see also \cite{siefert2020b} and references therein. 

% An interesting feature that we consistently observe in our FEM simulations is that the twist profiles are non-monotonic around each perversion. This is most easily seen in Fig.~\ref{fig:methodsFEM}c, which shows how the twist rate `overshoots' before it rapidly changes sign across perversions. This behavior is in contrast to the usual sigmoidal variation of the order parameter across phase boundaries, such as that predicted by the Cahn-Hilliard equation for phase separation in a binary fluid \citep{jones2002}. The observed overshoots are likely a consequence of the mechanism underlying torsional buckling, i.e., by twisting, the ribbon is able to stretch material further away from the centerline relative to material that is closer, thereby allowing it to effectively relieve compressive stress in the outer region without significantly adding to the tensile stress in the inner region. Because the twist rate passes through zero across perversions, this effect becomes diminished, for which the overshoot provides additional stretching to compensate. The effect is, therefore, analogous to the phenomenon of capillary-levelling in thin films, in which the free surface displays a similar overshoot near abrupt topographical features to relieve excess surface area \citep{mcgraw2011,mcgraw2012,bertin2021}.

We found that our FEM simulations are able to reproduce the main qualitative features of the experimental system (recall Figs.~\ref{fig:endshortsingleprestrain}a--b and \ref{fig:rotationsingleprestrain}a--b), despite differences in the precise buckling pattern. We attribute these differences to the presence of imperfections in the experiments, both in the ribbon samples and in the clamping conditions. Generally, we expect that the system becomes highly sensitive to imperfections near the onset of buckling. Because the incipient buckling mode is known to control the location of perversions in the related bi-strip studied by \cite{huang2012} and \cite{lestringant2017}, this sensitivity should persist throughout the entire loading history. The sensitivity is also exacerbated by the fact that perversions can move and interact with little energetic cost, as is easily observable experimentally by twisting or poking the ribbon by hand in the vicinity of perversions. As discussed in \S\ref{sec:ribbonmodel}, perversions are associated with higher-order gradient terms in the strain energy, which are asymptotically small compared to the energy associated with longitudinal stretching in the slender limit $h \ll a \ll \ell$. 
% Thus, in this limit, the re-arrangement of perversions can be regarded as a `zero stiffness' mode \citep{schenk2014} that requires negligible additional external work.

\subsection{Discussion and outlook}
\label{sec:conclusionsoutlook}

The phase diagram presented in Fig.~\ref{fig:ribbonvalidity}a provides insight into torsional instabilities observed in other elastic ribbons. While Fig.~\ref{fig:ribbonvalidity}a is only valid for the precise distribution of pre-stress considered here, i.e.~Eq.~\eqref{eqn:prestress}, we expect that a similar diagram is obtained whenever the pre-stress (with respect to a suitable reference configuration) is symmetric and more compressive towards the edges than around the centerline. For example, \cite{siefert2020a}, and \cite{gao2020} were able to program such a stress profile in a patterned fabric strip, which spontaneously adopted a helicoidal shape upon inflation. Similarly, in kelp blades, the growth rate is known to increase with distance from the blade center, and different buckling morphologies of kelp blades have been studied by \cite{koehl2008}. Narrow blades tend to buckle globally to helicoidal shapes, while wide blades buckle microscopically to form ruffled edges. This can be interpreted as moving between regions II and III on Fig.~\ref{fig:ribbonvalidity}a, noting that, for a fixed pre-strain $p$, increasing the width decreases $\eta$ and hence increases the re-scaled pre-strain $P$.

Our system can be useful in understanding torsional instabilities driven by residual stress in other prismatic solids. An example is the `twisters' studied by \cite{turcaud2011,turcaud2020}, which are rods composed of two elastic phases with contrasting expansion coefficients (when heated, for example). In numerical simulations, the rods were observed to spontaneously twist by an amount that depends sensitively on the cross-section geometry; in particular, \cite{turcaud2011} could only obtain significant twisting by connecting flattened `wings' of the high-expansion phase to a compact inner region composed of the low-expansion phase. Noting that the high and low-expansion phases are, respectively, analogous to the outer (initially unstressed) and inner (pre-stretched) regions in our ribbon system, this observation is consistent with the result of Eq.~\eqref{eqn:generaltwistvsbend}: twisting is only observed if the destabilizing effect of non-uniform pre-stress is sufficient to overcome the energy penalty of bending. Due to the flattened cross-section of the ribbon, the outer region naturally resembles the winged geometry of \cite{turcaud2011}.

Finally, our ribbon model can also be useful in developing a quantitative understanding of other torsional instabilities. The 1D energy density that we derived in Eq.~\eqref{eqn:ribbonenergydim} can, in principle, be generalized to other cross-section geometries. Provided that there is sufficient symmetry so that the straight, untwisted configuration is always an equilibrium solution, our analysis suggests that only uniaxial twisting and stretching about a straight centerline need to be considered. From this simple kinematic restriction, a similar energy may then readily be derived, which, in turn, should lead to a condition for instability in terms of the geometric parameters for a general cross-section shape. This approach opens up several directions for further study: for example, it can be used as the basis to design rods that become unstable at a pre-programmed threshold. When combined with other types of actuation such as inflation \citep{siefert2019,siefert2020a,jones2021,becker2022}, it can potentially be used to develop soft actuators whose deformation in three dimensions can be repeatedly re-programmed during deformation.

% \section*{Acknowledgments}
% We are grateful to 

\appendix
\gdef\thesection{\appendixname \Alph{section}} % corrected redefinition of "\thesection"
\makeatletter

\section{Relaxation of microscopic displacements in the ribbon model}
\label{sec:appendixrelaxation}

In this Appendix, we detail the relaxation procedure used to eliminate the dependence of the strain energy $\cW$ on the microscopic displacements $(u,v,w)$, and hence obtain the reduced energy density $\cW = \cW(\epsilon,k)$ reported in Eq.~\eqref{eqn:ribbonenergydim}. Our analysis follows Appendix A.2 in \cite{audoly2021b}, which we modify by (i) incorporating a uniaxial pre-stress $\nssref(T)$ and (ii) neglecting centerline bending. We note that the calculation below holds for any distribution of pre-stress, $\nssref(T)$.

Starting with the strain energy (per unit length) $\cW$ defined in Eq.~\eqref{eqn:strainenergy}, we impose the kinematic constraints in Eqs.~\eqref{eqn:kinematiccentroid}--\eqref{eqn:kinematictwistangle} by introducing the four scalar Lagrange multipliers $(\lambda_u,\lambda_v,\lambda_w,\mu_w)$ and forming the augmented energy:
\beqn
\cL\left(u,v,w;\epsilon,k,\nssref\right) \equiv \cW\left(u,v,w;\epsilon,k,\nssref\right) - \intwidth{\left[\lambda_u u(T) + \lambda_v v(T) + \left(\lambda_w + \mu_w T\right) w(T)\right]}.
\eeqn
Inserting the expression in Eq.~\eqref{eqn:strainenergy} for $\cW$ gives
\beqn
\cL\left(u,v,w;\epsilon,k,\nssref\right) \equiv \intwidth{\cF(T,u,v,w;\epsilon,k,\nssref)},
\eeqn
where we define
\beqa
\cF\left(T,u,v,w;\epsilon,k,\nssref\right) & \equiv & \frac{1}{2}\left(\Eab\Aabapbp\Eapbp + \frac{h^2}{12}\Bab\Aabapbp\Bapbp\right) + \nssref\Ess - \lambda_u u - \lambda_v v - \left(\lambda_w + \mu_w T\right)w, \nonumber \\
 & = & \frac{Y h}{2\left(1-\nu^2\right)}\left[\left(\Ess+\Ett\right)^2-2(1-\nu)\left(\Ess\Ett-\Est^2\right)\right] \nonumber \\
 && + \: \frac{Y h^3}{24\left(1-\nu^2\right)}\left[\left(\Bss+\Btt\right)^2-2(1-\nu)\left(\Bss\Btt-\Bst^2\right)\right] \nonumber \\
 && + \: \nssref\Ess - \lambda_u u - \lambda_v v - \left(\lambda_w + \mu_w T\right)w. \label{eqn:defncF}
\eeqa
Substituting the expressions in Eq.~\eqref{eqn:strainsdefmhomogeneous} for the membrane and bending strains, $\cF$ can be written as an explicit
function of $u$, $v$, $w$, $\epsilon$ and $k$; however, it is convenient to keep the strains unevaluated for now.

We then perturb the displacements $(u,v,w) \to (u +\delta u,v +\delta v,w +\delta w)$ and compute the first variation $\cF \to \cF + \delta \cF$. Because the membrane and bending strains depend only on the first derivatives of $u$ and $v$, the first variation is
\beqn
\delta \cF = -\lambda_u\delta u -\lambda_v\delta v + \pd{\cF}{w}\delta w + \pd{\cF}{u'}\delta u' + \pd{\cF}{v'}\delta v' + \pd{\cF}{w'}\delta w' + \pd{\cF}{w''}\delta w'',
\eeqn
where
\beqa
&& \pd{\cF}{w} = -\frac{Y h}{1+\nu}k\Est -\lambda_w - \mu_w T, \qquad 
\pd{\cF}{u'} = \frac{Y h}{1-\nu^2}\left(\Ett+\nu\Ess\right), \qquad
\pd{\cF}{v'} = \frac{Y h}{1+\nu}\Est, \nonumber \\
&& \pd{\cF}{w'} = \frac{Y h}{1-\nu^2}w'\left(\Ett+\nu\Ess\right) + \frac{Y h}{1+\nu}kT\Est, \qquad
\pd{\cF}{w''} = \frac{Y h^3}{12\left(1-\nu^2\right)}\left(\Btt+\nu\Bss\right). \label{eqn:partialderivsF} 
\eeqa

% To evaluate the partial derivatives above, we note from \eqref{eqn:strainsdefmhomogeneous} that the only non-zero derivatives of the strain components with respect to $(u,v,w)$ (and their derivatives) are
% \beqn
% \pd{\Est}{w} = -\frac{k}{2}, \quad \pd{\Est}{v'} = \frac{1}{2}, \quad \pd{\Est}{w'} = \frac{k}{2}T, \quad \pd{\Ett}{u'} = 1, \quad \pd{\Ett}{w'} = w', \quad \pd{\Btt}{w''} = 1.
% \eeqn
% Hence, from Eq.~\eqref{eqn:defncF}, we calculate

Integrating by parts over the width $(-a/2,a/2)$ to remove all derivatives on the perturbed quantities $(\delta u,\delta v,\delta w)$, the first variation of the functional $\cL$ is
\beqa
\delta \cL = \intwidth{\delta \cF} & = & \left\lbrace\pd{\cF}{u'}\delta u + \pd{\cF}{v'}\delta v + \left[\pd{\cF}{w'} - \left(\pd{\cF}{w''}\right)'\right]\delta w + \pd{\cF}{w''}\delta w'\right\rbrace\Bigg\lvert_{-a/2}^{a/2} \nonumber \\ 
&& \: - \intwidth{\left\lbrace\left[\lambda_u +\left(\pd{\cF}{u'}\right)'\right]\delta u + \left[\lambda_v +\left(\pd{\cF}{v'}\right)'\right]\delta v + \left[-\pd{\cF}{w} + \left(\pd{\cF}{w'}\right)' - \left(\pd{\cF}{w''}\right)''\right]\delta w\right\rbrace}. \qquad \quad \label{eqn:variationL}
\eeqa
Requiring that the first variation $\delta \cL$ is zero for all admissible $(\delta u,\delta v,\delta w)$, we obtain the Euler-Lagrange equations:
\beq
\lambda_u +\left(\pd{\cF}{u'}\right)' = 0, \quad
\lambda_v +\left(\pd{\cF}{v'}\right)' = 0, \quad  -\pd{\cF}{w} + \left(\pd{\cF}{w'}\right)' - \left(\pd{\cF}{w''}\right)'' = 0, \qquad T \in \left(-\frac{a}{2},\frac{a}{2}\right), \label{eqn:ELeqns}
\eeq
and the natural boundary conditions
\beq
\pd{\cF}{u'} = \pd{\cF}{v'} = \pd{\cF}{w'} - \left(\pd{\cF}{w''}\right)' = \pd{\cF}{w''} = 0, \qquad T = \pm \frac{a}{2}. \label{eqn:natBCs}
\eeq
Using the expressions in Eq.~\eqref{eqn:partialderivsF}, these simplify to
\beqn
\Ett + \nu\Ess = \Est = \Btt + \nu\Bss = \left(\Btt + \nu\Bss\right)' = 0, \qquad T = \pm \frac{a}{2}.
\eeqn
These correspond to zero transverse membrane stress, shear stress, transverse bending stress and shear force at the ribbon boundaries. 

It is possible to reduce Eqs.~\eqref{eqn:ELeqns}--\eqref{eqn:natBCs} to a boundary-value problem for the out-of-plane displacement, $w$. To this end, we integrate the first two Euler-Lagrange equations across the width, and make use of the first two natural boundary conditions in Eq.~\eqref{eqn:natBCs} to yield
\beqn
\lambda_u = \lambda_v = 0.
\eeqn
Returning to Eq.~\eqref{eqn:ELeqns}, we see that $\partial\cF/\partial u'$ and $\partial\cF/\partial v'$ are constant. Again using the boundary conditions in Eq.~\eqref{eqn:natBCs}, we obtain
\beqn
\pd{\cF}{u'} = 0, \quad \pd{\cF}{v'} = 0, \qquad T \in \left(-\frac{a}{2},\frac{a}{2}\right),
\eeqn
i.e., the transverse membrane stress $(\Ett+\nu\Ess)$ and shear stress $\Est$ are everywhere zero. The remaining partial derivatives in Eq.~\eqref{eqn:partialderivsF} simplify to
\beqn
\pd{\cF}{w} = -\lambda_w - \mu_w T, \qquad 
\pd{\cF}{w'} = 0, \qquad
\pd{\cF}{w''} = \frac{Y h^3}{12\left(1-\nu^2\right)}\left(\Btt+\nu\Bss\right).
\eeqn
Substituting for $\Btt$ and $\Bss$ using Eq.~\eqref{eqn:strainsdefmhomogeneous}, the final Euler-Lagrange equation in Eq.~\eqref{eqn:ELeqns} becomes
\beq
\lambda_w + \mu_w T - \frac{Y h^3}{12\left(1-\nu^2\right)}w'''' = 0, \label{eqn:fullODEw}
\eeq
to be solved with the remaining natural boundary conditions in Eq.~\eqref{eqn:natBCs} and the kinematic conditions for $w$ in Eqs.~\eqref{eqn:kinematiccentroid}--\eqref{eqn:kinematictwistangle}:
\beq
 w'' = w''' = 0 \quad \mathrm{at} \quad T = \pm \frac{a}{2}, \qquad \intwidth{w} = \intwidth{T w} = 0. \label{eqn:fullCsw}
\eeq

Furthermore, we can eliminate the Lagrange multipliers $\lambda_w$ and $\mu_w$ as follows. Integrating Eq.~\eqref{eqn:fullODEw} over the width and using $w'''(\pm a/2) = 0$ gives
\beqn
\lambda_w = 0.
\eeqn
Similarly, if we multiply Eq.~\eqref{eqn:fullODEw} by $T$ before integrating (using parts to integrate the final term), the boundary conditions $w''(\pm a/2) = w'''(\pm a/2) = 0$ give
\beqn
\mu_w = 0.
\eeqn
Returning to Eqs.~\eqref{eqn:fullODEw}--\eqref{eqn:fullCsw}, the boundary-value problem for $w$ reduces to
\beqan
&& \qquad \qquad \qquad \quad w'''' = 0,  \qquad T \in \left(-\frac{a}{2},\frac{a}{2}\right), \\
&& w'' = w''' = 0 \quad \mathrm{at} \quad T = \pm \frac{a}{2}, \qquad \intwidth{w} = \intwidth{T w} = 0.
\eeqan
The unique solution is simply $w(T) = 0$; the equations $\Ett+\nu\Ess = \Est = 0$ can then be integrated with the other kinematic conditions in Eqs.~\eqref{eqn:kinematiccentroid}--\eqref{eqn:kinematictwistangle} to determine the in-plane displacements as $u(T) = -\nu(\epsilon T + k^2T^3/6)$ and $v(T) = 0$.

In summary, we have shown that, as a result of relaxation of the microscopic displacements with respect to the macroscopic strains $(\epsilon,k)$, we have
\beqn
\lambda_u = \lambda_v = \lambda_w = \mu_w = 0, \quad u(T) = -\nu\left(\epsilon T + \frac{k^2}{6}T^3\right), \quad v(T) = w(T) = 0, \qquad T \in \left(-\frac{a}{2},\frac{a}{2}\right).
\eeqn
The strains in Eq.~\eqref{eqn:strainsdefmhomogeneous} associated with homogeneous solutions then reduce to
\beq
\Ess(T) = \epsilon  + \frac{k^2}{2}T^2, \quad \Est(T) = 0, \quad \Ett(T) = -\nu\Ess(T), \qquad \Bss(T) = 0, \quad \Bst(T) = k, \quad \Btt(T) = 0. \label{eqn:strainsrelaxed}
\eeq
The only non-zero components that remain are (i) the longitudinal strain $\Ess$ resulting from the macroscopic strain $\epsilon$ and the stretching that arises due to twist; (ii) the transverse in-plane strain $\Ett$ that arises due to Poisson effects; and (iii) the shear strain $\Bst$ associated with twisting.

With the above expressions, the energy function $\cF$ in Eq.~\eqref{eqn:defncF} simplifies to
\beqn
\cF = \frac{Y h}{2}\left\lbrace\epsilon^2 + \left[\epsilon T^2 +\frac{h^2}{6(1+\nu)}\right]k^2 + \frac{T^4}{4}k^4\right\rbrace + \nssref\left(\epsilon + \frac{T^2}{2}k^2\right).
\eeqn
Integrating this across the width yields the expression reported in Eq.~\eqref{eqn:ribbonenergydim}.

\section{Determining the pre-stress using the neo-Hookean material model}
\label{sec:appendixprestress}

To implement the pre-stress in the framework of the neo-Hookean material model, we identify the principal direction $3$ with the longitudinal direction along the fully-unrelaxed ribbon, and the $1$, $2$ directions with the transverse directions. Under the applied pre-strain $p$, the principal stretches in the inner region are then
\beq
\lambda_3 = \lambda_u \equiv 1 + p, \quad \lambda_1 = \lambda_2 = \left(\frac{J}{\lambda_u}\right)^{1/2}, \label{eqn:principalstretches}
\eeq
where the volume ratio $J$ is to be determined. The Cauchy (true) stresses are derived from the strain energy in Eq.~\eqref{eqn:strainenergyneohook} as
\beq
\sigma_i = \frac{\lambda_i}{J}\pd{U}{\lambda_i} \quad (i = 1,2,3). \label{eqn:cauchystresses}
\eeq
After substituting the principal stretches \eqref{eqn:principalstretches}, we obtain
\beq
\sigma_3 = \frac{2}{3}\mu J^{-5/3}\left(\lambda_u^2 - \frac{J}{\lambda_u}\right) + K_b(J-1), \quad
\sigma_1 = \sigma_2 = -\frac{1}{3}\mu J^{-5/3}\left(\lambda_u^2 - \frac{J}{\lambda_u}\right) + K_b(J-1).
\label{eqn:cauchystressesuniaxial}
\eeq
Because the pre-stretched strip is initially unconstrained transversely (Fig.~\ref{fig:problemdefn}b), we have $\sigma_1 = \sigma_2 = 0$, yielding:
\beq
-\frac{1}{3}\mu J^{-5/3}\left(\lambda_u^2 - \frac{J}{\lambda_u}\right) + K_b(J-1) = 0. \label{eqn:Jeqnneohook}
\eeq
Once this equation is solved (using a standard root-finding algorithm) for $J(\lambda_u)$, the pre-stress $\sigma_3 = \sigma_0$ can be evaluated using Eq.~\eqref{eqn:cauchystressesuniaxial}. 

We note that in the small-strain limit $p \ll 1$, ignoring terms of $O(p^2)$, we have $J \sim 1 + (1-2\nu)p$ and $\sigma_0 \sim Y p$ (making use of the expressions $\mu = Y/[2(1+\nu)]$ and $K_b = Y/[3(1-2\nu)]$), thus recovering the expression for $\sigma_0$ used in the ribbon model (Eq.~\eqref{eqn:prestress}).

% \beq
% \sigma_0 = \mu J^{-5/3}\left(\lambda_u^2 - \frac{J}{\lambda_u}\right). \label{eqn:prestressneohook}
% \eeq

\section{Determining the bending and twisting strains along the ribbon centerline}
\label{sec:appendixpostprocess}

Here, we describe our method to determine the bending and twisting strains along the ribbon centerline, for each loading step in our numerical (FEM) simulations. Our approach computes the twist angle by averaging data from all mid-surface nodes in a given cross-section, accounting for possible bending of the centerline (associated with rotation of the unit tangent vector). It is therefore robust to warping of the cross-section and distortion of mesh elements. We found that characterizing the twist using a single off-centerline node often gave noisy and inaccurate results.

Starting from the raw data of node positions along the ribbon mid-surface, we first determined the orthonormal director basis $\lbrace\bd_1,\bd_2,\bd_3\rbrace$; here $\bd_3$ is the unit tangent vector (we assumed $\bd_3 = \be_z$ in the ribbon model derived in \S\ref{sec:ribbonmodel}), and $\bd_1$ and $\bd_2$ span the cross-section perpendicular to $\bd_3$. Firstly, we obtained the coordinates of all nodes on the centerline, sorted in order of increasing material coordinate $S$ (the existence of node points exactly on the material centerline was guaranteed from the symmetry of the mesh). Using centered finite differences on the interior centerline nodes, we then calculated the unit tangent vector, $\bd_3$ (for the edge nodes, the tangent was unchanged from $\be_z$ due to the clamped boundary conditions). Next, we calculated the directors $\bd_1$ and $\bd_2$ at each $S$ using the coordinate data of all mid-surface nodes in that cross-section. Specifically, we first computed the displacement vectors from the deformed centerline to all mid-surface nodes (components with respect to the global Cartesian frame), which were projected into the plane normal to $\bd_3$. This enabled us to calculate `trial' directors $\bd_1$ and $\bd_2$, for example by aligning $\bd_1$ with the displacement vector to the nearest off-centerline node, and setting $\bd_2 = \bd_3 \times \bd_1$. We then computed the plane-polar coordinates of the projected nodes in the trial $\lbrace\bd_1,\bd_2\rbrace$ frame, and defined the average twist angle of the cross-section to be the second (polar) moment of the angular coordinates. This definition is the discrete analog of the kinematic condition (recall Eq.~\eqref{eqn:kinematictwistangle}) used to uniquely define the twist angle in the ribbon model. Finally, the directors $\bd_1$ and $\bd_2$ were found by rotating the trial directors such that $\bd_1$ aligned with the average twist angle.

In general, under the requirement of orthonormality, the evolution of the directors along the centerline (parameterized by the material coordinate $S$) is described by the Darboux vector $\bka(S)$ where
\beq
\bd_i'(S) = \bka(S)\times\bd_i(S), \qquad i = 1,2,3. \label{eqn:defndarboux}
\eeq
Writing $\bka(S)$ in terms of components in the director basis, i.e., $\bka(S) = \kappa_1(S)\bd_1(S) + \kappa_2(S)\bd_2(S) + \kappa_3(S)\bd_3(S)$,
% \beq
% \bka(S) = \kappa_1(S)\bd_1(S) + \kappa_2(S)\bd_2(S) + \kappa_3(S)\bd_3(S), \label{eqn:expanddarboux}
% \eeq
we can interpret $\kappa_1(S)$, $\kappa_2(S)$ as the bending strains and $\kappa_3(S) = k(S)$ as the twisting strain (the rate of twist along $S$). Thus, with the directors $\lbrace\bd_1,\bd_2,\bd_3\rbrace$ determined on the numerical mesh, the bending and twisting strains were calculated using the following equations, which immediately follow from Eq.~\eqref{eqn:defndarboux}:
\beqn
\kappa_1(S) = -\bd_3'(S)\cdot\bd_2(S), \quad
\kappa_2(S) = \bd_3'(S)\cdot\bd_1(S), \quad
\kappa_3(S) = \bd_1'(S)\cdot\bd_2(S).
\eeqn
In these expressions, the derivatives were evaluated using centered finite differences on the numerical mesh.

%\bibliographystyle{elsarticle-harv} 
%\bibliography{Prestressed_Ribbons_Refs}


\end{document}